\documentclass[11pt]{article}

\usepackage{subfigure,epsf,graphicx,amsfonts,amsmath,amssymb}
\usepackage[USenglish]{babel}
\usepackage{booktabs}
\usepackage{soul}
\usepackage[latin1]{inputenc}
\usepackage{csquotes}
\usepackage{physics}
\usepackage{hyperref}
\usepackage{url}
\usepackage[capitalize]{cleveref}

\newcommand{\field}[1]{\mathbb{#1}}
\newcommand{\N}{\field{N}}

\newcommand{\ffrac}[2]{\frac{\displaystyle #1}{\displaystyle #2}}

\usepackage{subfigure}

\usepackage{wrapfig}

\newcommand{\CR}[2]{%
\rotatebox[origin=c]{90}{R#2} 
\left\{
\begin{array}{rcl}
\dot{x}_{#2} & = & - y_#2 - z_#2  #1 \\
\dot{y}_{#2} & = &  x_{#2} + a_#2 \, y_{#2}  \\ 
\dot{z}_{#2} & = & b_#2 + c_#2 z_{#2} + x_{#2} z_#2 
\end{array} 
\right. %
}

\begin{document}

{\huge \bf Causal Dynamic Resonance\footnote{accepted for publication in PNAS}}

    Claudia~Lainscsek$^{1,2}$, Pariya Salami$^{3,4}$, Simon Dr\"ager$^{1}$, Adi~R.~Bulsara$^{5}$, Sydney~S.~Cash$^{3,4}$, and Terrence~J.~Sejnowski$^{1,2,6}$

$^1$~Computational Neurobiology Laboratory, The Salk Institute for Biological Studies, 10010 North Torrey Pines Road,
La Jolla, CA 92037, USA \\
$^2$~Institute for Neural Computation, University of California San Diego, La Jolla, CA 92093, USA \\
$^3$~Department of Neurology, Massachusetts General Hospital and Harvard Medical School, Boston, MA 02114, USA\\
$^4$~Center for Neurotechnology and Neurorecovery, Department of Neurology, Massachusetts General Hospital, Boston, MA, USA \\
$^5$~Naval Information Warfare Center (RET), San Diego, CA 92152, USA \\
$^6$~Department of Neurobiology, University of California San Diego, La Jolla, CA 92093, USA

\begin{abstract}%
    \noindent%
Many dynamical systems in nature, such as brains and weather systems, are highly nonlinear and complex.  Determining information flow among the components that make up these dynamical systems is challenging. If the components are the result of a common process or become synchronized, causality measures typically fail.  We previously introduced Cross-Dynamical Delay Differential Analysis (CD-DDA), a nonlinear method for assessing causal influence, along with a complementary approach for  dynamical similarity between time series data, Dynamical Ergodicity Delay Differential Analysis (DE-DDA). Here, we show that ``Causal Dynamic Resonance (CDR)'' further improves the false positive rejection rate by adding white noise to the data, without perturbing the underlying dynamical system. This is followed by a study of CDR in coupled R\"ossler systems, where ground truth interactions are known and in invasive intracranial electroencephalographic (iEEG) data from drug-resistant epilepsy patients undergoing presurgical monitoring.
\end{abstract}



\begin{wrapfigure}{R}{0.4\textwidth}
    \centering
    \includegraphics[width=0.4\columnwidth]{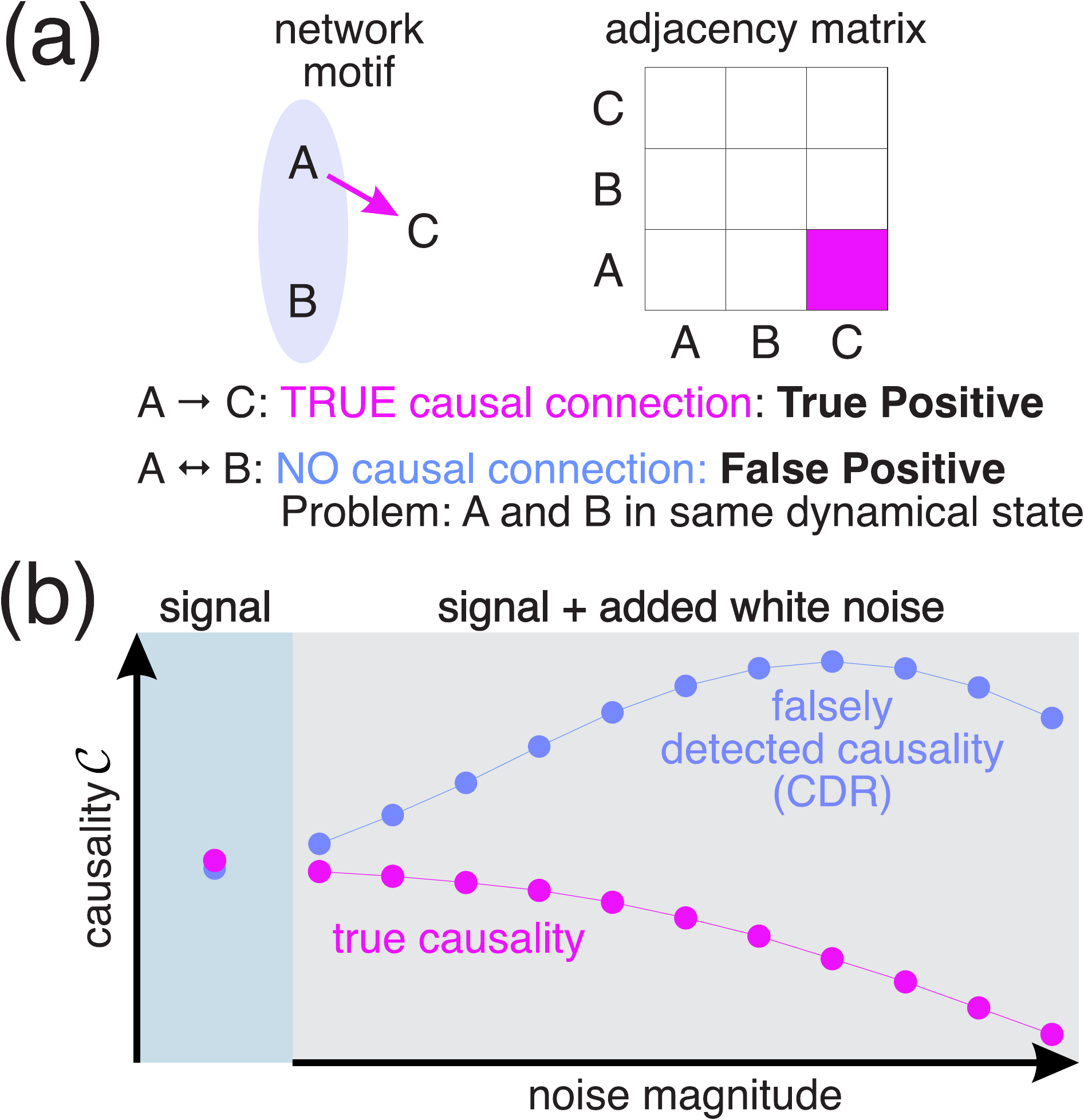}
	\hspace*{-1cm}
    \caption{\textbf{Causal Dynamic Resonance (CDR)}. 
    (a): Network motif and adjacency matrix: nodes A and B are in a similar dynamical state with no causal connection. Nodes A and C are connected through a causal connection.
    (b): The CD-DDA (causality) outputs from two signals (magenta and violet dots) are shown in the blue box. 
    The effect of added white noise is shown in the gray box.
    Adding white noise to the magenta signal decreases the magnitude of the causality output. For the violet signal on the other hand additive white noise increases the magnitude of the causality output and decreases it for even larger noise intensities. The violet curve shows CDR, while the magenta curve does not.     }
    \label{fig:SR_explain}
\end{wrapfigure}

\textbf{Significance:}
A major challenge in neuroscience is distinguishing true causal interactions from spurious connections in complex neural networks. Here, we introduce a nonlinear dynamical analysis framework that accurately identifies causal relationships while eliminating falsely inferred connections. Validation in a well- characterized dynamical system demonstrates its ability to recover true network interactions, providing a robust approach for studying information flow in the brain and other complex systems.

\section{Introduction}

Causality, synchronization, and dynamical similarity in highly complex nonlinear systems like brains or weather systems are distinct but related by the unknown deterministic structure of the underlying dynamical system. Therefore, distinguishing them is challenging.  For two systems that are not independent of each other, either because they result from a common process or they are already synchronized, causality measures typically fail.  In~\cite{lainscsek19-101103} we introduced Cross-Dynamical Delay Differential Analysis (CD-DDA) to assess causality and in~\cite{lainscsek23-123136} we extended this framework to network-level analyses.  In~\cite{lainscsek21-103108} we incorporated Dynamical Ergodicity 
DDA (DE-DDA), a measure of dynamical similarity between time series, to eliminate some falsely identified causal connections. Here, we 
introduce \textbf{causal dynamic resonance (CDR)} that further improves the false
positive rejection rate by adding white noise to the signals.

\Cref{fig:SR_explain} shows an example of a network motif with three nodes.  Nodes A and B are not causally connected but in a similar dynamical state (e.g.\ same brain state) while C is different. CD-DDA as well as other causality measures typically identify A~$\leftrightarrow$~B as well as A~$\rightarrow$~C as causal connections.  CDR aims to distinguish between falsely detected causal connections A~$\leftrightarrow$~B and true connections A~$\rightarrow$~C. Adding Gaussian white noise with a varying signal-to-noise ratio (SNR) changes the CD-DDA outputs dramatically: While it generally goes down for the true connection A~$\rightarrow$~C, it first increases and then decreases for the false connections A~$\leftrightarrow$~B as indicated in \cref{fig:SR_explain}. We evaluate and demonstrate this effect across a range of network configurations using simulated data from coupled R\"ossler systems where ground truth is known as well as for iEEG data from epilepsy patients. 

\subsection{Causality}

As pointed out by Yule in 1926 \cite{yule26-1},  correlation does not imply causation. Yule also made a connection between the introduction of delays and causal relations between time series.  In 1969 Granger \cite{granger69-424}  introduced a statistical measure of causality that is widely used in signal processing. This work is closely related to the work of  Wiener, which was published in 1956 \cite{wiener56}.
Since  Granger causality (GC) relies on linear autoregressive models, it may not yield good results for some nonlinear systems.   To circumvent the limitations of the linear Granger causality test, Brock et al.\ \cite{brock87, brock96} proposed a test based on correlation integrals \cite{grassberger83-189} and Baek and Brock \cite{baek92}, Hiemstra and Jones \cite{hiemstra94-1639}, and Bai \cite{bai10-5, bai18-0185155} then introduced nonlinear Granger causality.
In 2000 Schreiber \cite{schreiber00-461} introduced transfer entropy (TE) for information transfer  between nonlinear dynamical systems. If the systems are linear Gaussian processes, GC and TE are equivalent \cite{barnett09-238701}.
Determining causality from dynamical attractors of nonlinear dynamical systems and the concept of generalized synchronization were introduced by 
Schiff \cite{schiff96-6708}, 
Arnhold \cite{arnhold99-419}, 
Hirata \cite{hirata10-016203, hirata16-e0158572}, and
Sugihara \cite{sugihara12-496}, 
among others.
One implementation is Sugihara's convergent cross mapping (CCM)  \cite{sugihara12-496}.  
CCM is based on standard uniform delay embeddings and is therefore limited to a subset of the dynamical systems found in nature.  In 2018 a focus issue in Chaos was published; it summarized recent developments for causality detection \cite{bollt18-075201}. Recent articles use complex network theory to characterize causality of multivariate data \cite{zou19-1, ruan19-04311}.
Adding noise to the data has been used to identify true causal connections based on the idea that
real causal relationships are robust, while spurious correlations collapse.
Such methods include invariant causal prediction \cite{peters16-947},
structural causal models \cite{pearl09}, and
additive noise models \cite{hoyer08}. In these approaches false causal connections disappear with added noise while in CDR false positives increase and then decrease when noise is added to the data. 

\subsection{Delay Differential Analysis (DDA)}
DDA is a detection and classification framework that combines differential embeddings with linear and nonlinear nonuniform functional delay embeddings~\cite{takens81-366,packard80-712,sauer91-579} to relate the current derivatives of a system to the 
past values of the system variables~\cite{kremliovsky97-57, lainscsek13-182, lainscsek19-101103}.

More traditional linear analyses often characterize each data segment by numerous quantitative descriptors, hereafter referred to as \emph{features}, many of which are derived from spectral content. This can amount to hundreds of features per segment, and artificial neural network-based approaches may enlarge the representation further. As a result, dimensionality reduction is typically required to make the analysis tractable.

Nonlinear systems can be represented by Volterra series \cite[]{Volterra:30,Schetzen:89,Rugh:81}. 
Such series are characterized by an expansion of the form
\begin{equation}
\label{eq:volt}
y(t) = \sum_{n=1}^{\infty}~ \int\limits_{-\infty}^{-\infty} \ldots \int\limits_{-\infty}^{-\infty} ~
	h_n(\tau_1, \tau_2, \ldots \tau_n)~\prod_{i=1}^n x(t-\tau_i) ~d\tau_i
\end{equation}
where $x(t)$ is the input, $y(t)$ is the output and $h(\star)$ is the Volterra kernel.
The Volterra series expansion is the exact analog of the expansion of a nonlinear function as a power series.
Such an expansion may present numerical difficulties because
higher order terms are almost linearly dependent. To overcome these difficulties, the Wiener
series  was introduced \cite{Wiener:58}. Wiener series can be regarded as a transformed version
of the Volterra series whose terms are orthogonal to each other,
provided the input is white Gaussian. The corresponding kernels are called the Wiener kernels.
The specific problem when identifying a nonlinear system using the Volterra series is to
determine the Volterra (Wiener) kernels from input-output time data which results in a high number of model parameters.
Interesting ideas to avoid the explosion of the
number of Volterra parameters are e.g.\ Volterra
kernel expansions using orthonormal Laguerre functions \cite{Dumont_Fu:92} or
using the Kronecker product structure for finite Volterra series \cite{Nowak_Veen:94c}.
DDA is loosely based on the discrete Volterra model,
where we have $\dot{y}$ on the left side of Eq.~[\ref{eq:volt}] and integrals are replaced by sums.
This then can be simplified to the general nonlinear DDA model
\begin{equation}
\label{GeneralDDA}
\dot{u}  = \sum\limits_{i=1}^{I} a_i \prod\limits_{n=1}^{N} u_{\tau_n}^{m_{n,i}} 
= \mathcal{F}_u
\end{equation}
for $\tau_n, m_{n,i} \in \N{}_0$,
where $N$ is the number of delays, $I$ is the number of terms, and $u_{\tau_n} =u(t - \tau_n )$. 
To restrict complexity of the model we use three-term models ({$I=3$}) with two delays ($N=2$) and an order of nonlinearity 
$\sum_i m_{n,i}\le4$.
Although one of the objectives of the Volterra series is to build a nonlinear model that captures not only the linear convolution but also the so-called higher order convolutions, DDA uses Eq.~[\ref{GeneralDDA}] to retrieve information on the underlying dynamics by looking at how the residuals generated from the regression problem behave as time passes by.  
We therefore use the coefficients $a_i$ and the least square error $\rho_u$  as features.

The first item in \cref{flavors} is \textbf{ST-DDA} (Single-Timeseries DDA). Here we apply DDA to the R\"ossler system~\cite{roessler76-397}. We used the same model
and delays as in~\cite{lainscsek19-101103} for the $x$-component of the R\"ossler system:
\begin{equation}
\label{DDA-model}
    \begin{array}{rcl}
    \dot{u} & = & a_1 \, u_1 + a_2 \, u_2 + a_3 \, u_1^3 = \mathcal{F}_u\\ 
    \end{array}
\end{equation}
with $u_j = u(t - \tau_j)$, $\tau_1 = 32~\delta t$,  $\tau_2 = 9~\delta t$, and $\delta t = 0.025$. We then apply the DDA model
equation~\eqref{DDA-model} to a data segment of length $L$:
\begin{equation}
\begin{array}{c}
\resizebox{0.88\columnwidth}{!}{$
\begin{array}{rcl}
\left(
\begin{array}{c}
\dot{u}(t+1) \\
\dot{u}(t+2) \\
\dot{u}(t+3) \\
\vdots \\
\dot{u}(t+L) \\
\end{array}
\right)
&=& 
\left(
\begin{array}{ccc}
u(t+1 - \tau_1) & u(t+1 - \tau_2)  & u(t+1 - \tau_1)^3  \\
u(t+2 - \tau_1) & u(t+2 - \tau_2)  & u(t+2 - \tau_1)^3  \\
u(t+3 - \tau_1) & u(t+3 - \tau_2)  & u(t+3 - \tau_1)^3  \\
& \vdots \\
u(t+L - \tau_1) & u(t+L - \tau_2)  & u(t+L - \tau_1)^3 
 \end{array}
\right) \,
\left(
\begin{array}{c}
  a_{1} \\
  a_{2} \\
  a_3
\end{array}
\right)
\end{array}
$}
\\ 
\\
{\bf \dot{u}} = {\bf M_u} \, {\bf A}
\end{array}
\label{data_SVD}
\end{equation}
to estimate the three coefficients \(a_1, a_2, a_3\) using singular value decomposition (SVD;~\cite{recipes}). 
The derivative on the left side is computed using 
the central derivative~\cite{miletics04-105, miletics05-1167}.
Then the root least square error 
\begin{equation}\label{rho}
    \rho_u = \sqrt{ \sum (\dot{u}  - \mathcal{F}_u)^2 }
\end{equation}
is computed, which we will be focusing on in this manuscript.
Note, that the discrete time $t$ in Eq.~[\ref{data_SVD}] must be greater than the largest delay plus the number of data points to compute the numerical derivative.

\begin{table}[htb]
\begin{center}
\begin{tabular}{c|c|c}
        Flavor & Reference & Description \\ \hline
        ST-DDA & \cite{lainscsek17-3181} & 
        \begin{minipage}[c][1cm]{0.6\textwidth}
        \includegraphics[width=\textwidth]{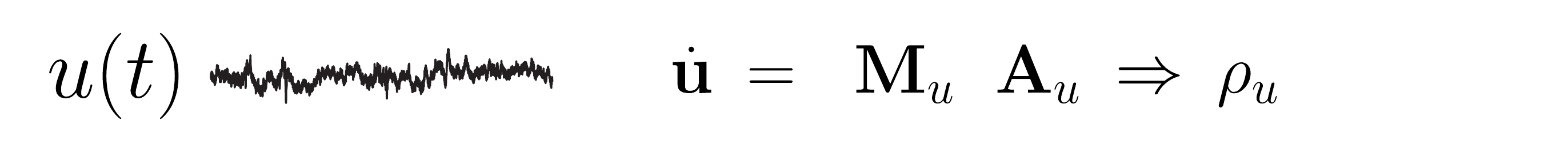} 
        \end{minipage}
        \\ \hline
        CT-DDA & \cite{Lainscsek19-3847} &
        \begin{minipage}[c][1.8cm]{0.6\textwidth}
        \includegraphics[width=\textwidth]{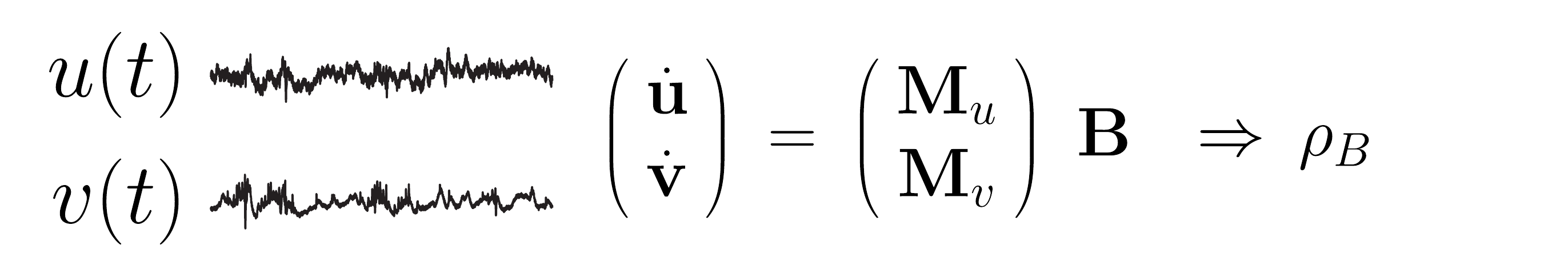} 
        \end{minipage}
        \\ \hline
        CD-DDA & \cite{lainscsek19-101103} &
        \begin{minipage}[c][4.5cm]{0.6\textwidth}
        \includegraphics[width=\textwidth]{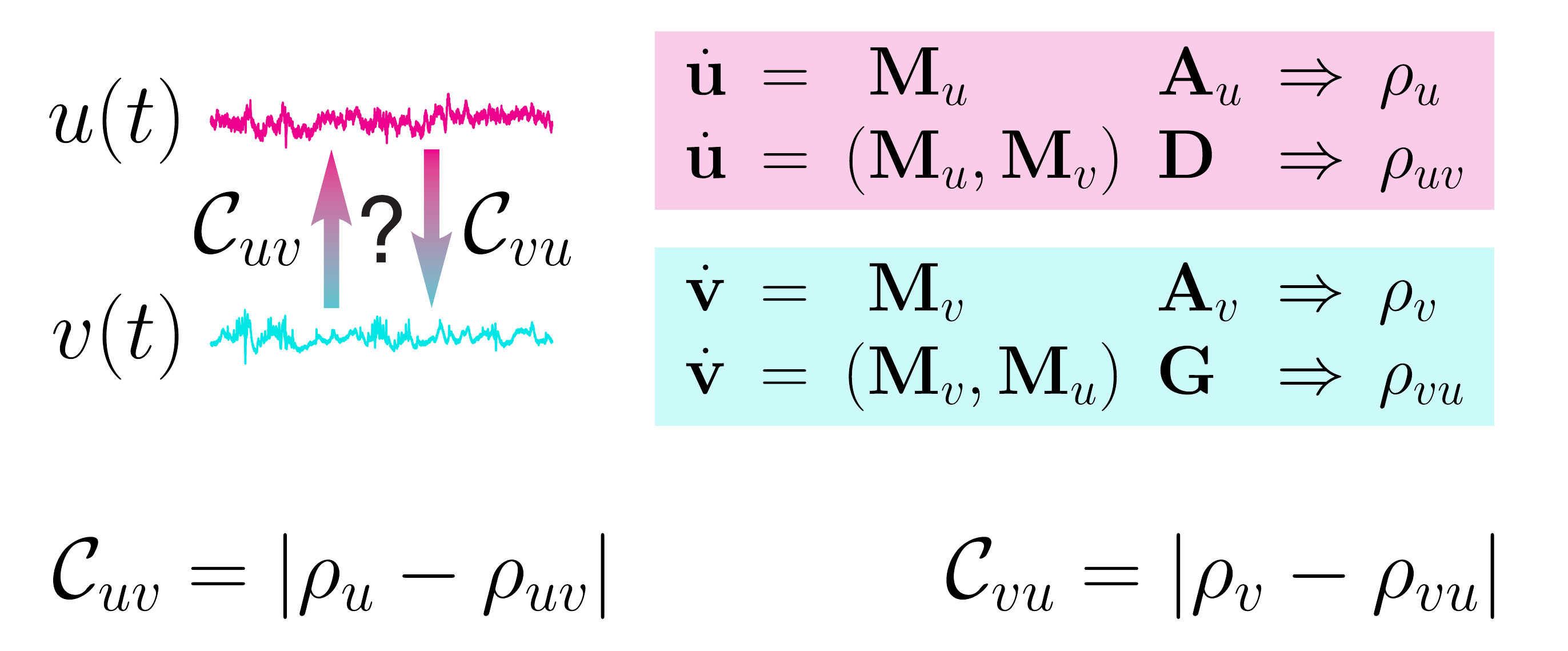} 
        \end{minipage}
        \\ \hline
        DE-DDA & \cite{lainscsek21-103108} &
        \begin{minipage}[c][2.4cm]{0.6\textwidth}
        \includegraphics[width=\textwidth]{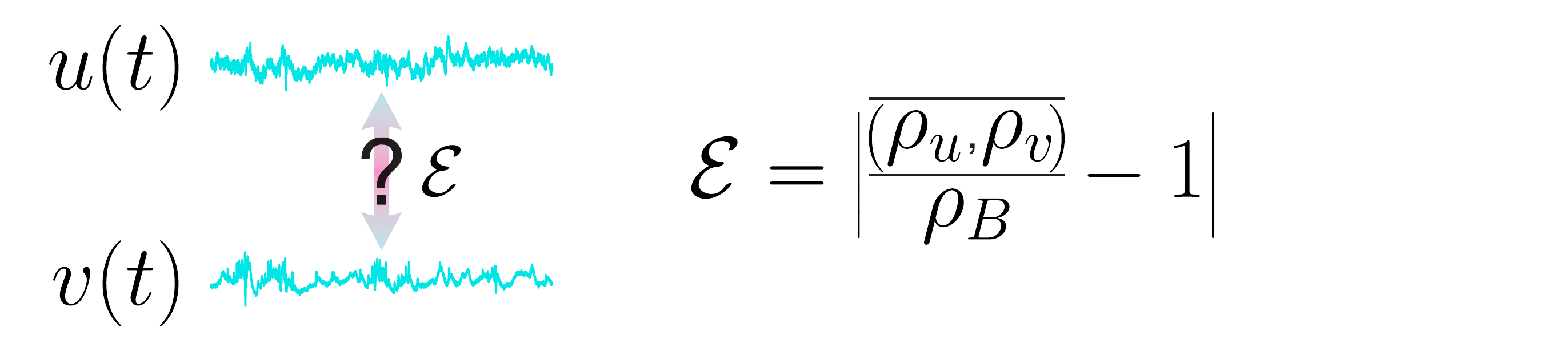} 
        \end{minipage}
\end{tabular}
\end{center}
\caption{\textbf{The flavors of DDA:} {\normalfont Single-Timeseries DDA (ST-DDA) is the classical variant developed for analyzing single time series.  Cross-Timeseries DDA (CT-DDA) determines the overall dynamics of multiple time series simultaneously. Cross-Dynamical DDA (CD-DDA)  measures causality between two time series. Dynamical-Ergodicity DDA (DE-DDA) is a combination of ST-DDA and CT-DDA to assess dynamical ergodicity or similarity from data.}
}    
\label{flavors}
\end{table}

While ST-DDA is the single-variate version for one timeseries, the other DDA flavors are multivariate versions on multiple timeseries. For
\textbf{CT-DDA} (Cross-Timeseries DDA) for example, we stack the derivatives of the timeseries as well as the delayed versions on top of
each other (\cref{flavors} shows the example of two time series) to estimate the coefficients from multiple timeseries simultaneously. This
modifies \eqref{data_SVD} for two timeseries $u(t)$ and $v(t)$ of length $L$ to
\begin{equation}
\begin{array}{c}
\resizebox{0.88\columnwidth}{!}{$
\begin{array}{rcl}
\left(
\begin{array}{c}
\dot{u}(t+1) \\
\dot{u}(t+2) \\
\dot{u}(t+3) \\
\vdots \\
\dot{u}(t+L) \\ [1ex]
\dot{v}(t+1) \\
\dot{v}(t+2) \\
\dot{v}(t+3) \\
\vdots \\
\dot{v}(t+L) 
\end{array}
\right)
&=& 
\left(
\begin{array}{ccc}
u(t+1 - \tau_1) & u(t+1 - \tau_2)  & u(t+1 - \tau_1)^3  \\
u(t+2 - \tau_1) & u(t+2 - \tau_2)  & u(t+2 - \tau_1)^3  \\
u(t+3 - \tau_1) & u(t+3 - \tau_2)  & u(t+3 - \tau_1)^3  \\
& \vdots \\
u(t+L - \tau_1) & u(t+L - \tau_2)  & u(t+L - \tau_1)^3 \\ [1ex]
v(t+1 - \tau_1) & v(t+1 - \tau_2)  & v(t+1 - \tau_1)^3  \\
v(t+2 - \tau_1) & v(t+2 - \tau_2)  & v(t+2 - \tau_1)^3  \\
v(t+3 - \tau_1) & v(t+3 - \tau_2)  & v(t+3 - \tau_1)^3  \\
& \vdots \\
v(t+L - \tau_1) & v(t+L - \tau_2)  & v(t+L - \tau_1)^3 
 \end{array}
\right) \,
\left(
\begin{array}{c}
  b_{1} \\
  b_{2} \\
  b_3
\end{array}
\right)
\end{array}
$}
\\ 
\\
\begin{array}{c}
\begin{array}{rcl}
\left(
\begin{array}{c}
{\bf \dot{u}} \\
{\bf \dot{v}} \\
\end{array}
\right)
& = &
\left(
\begin{array}{c}
{\bf M_u} \\
{\bf M_v} 
\end{array}
\right)
\, {\bf B}
\end{array}
\end{array}
\end{array}
\label{eq:CT-DDA}
\end{equation}
Thus, we have twice as much data to estimate the coefficients $(b_1, b_2, b_3)$. The number of stackable time series is arbitrary.
The only restriction is that all time series are dynamically similar. 

To test for dynamical similarity of two time series $u(t)$ and $v(t)$
we use \textbf{DE-DDA} (Dynamical Ergodicity DDA) \cite{lainscsek21-103108}  where we compare the mean of the errors of ST-DDA with the combined error of CT-DDA:
We compute the ST-DDA and CT-DDA errors 
\begin{equation}
\begin{array}{lll}
& {\bf \dot{u}} = {\bf M_u} \, {\bf A_u} & \Rightarrow \rho_u \\
& {\bf \dot{v}} = {\bf M_v} \, {\bf A_v} & \Rightarrow \rho_{v} \\
\text{and} & & \\
& 
\begin{array}{rcl}
\left(
\begin{array}{c}
{\bf \dot{u}} \\
{\bf \dot{v}} \\
\end{array}
\right)
& = &
\left(
\begin{array}{c}
{\bf M_u} \\
{\bf M_v} 
\end{array}
\right)
\, {\bf B}
\end{array}  & \Rightarrow \rho_{B} 
\end{array}
\end{equation}
If the two time series $u(t)$ and $v(t)$ are dynamically similar, then the errors \(\rho_u\) and \(\rho_v\) should be approximately equal. We therefore
use dynamical ergodicity
\begin{equation}
\mathcal{E} = \left| \ffrac{\overline{(\rho_u , \rho_v)}}{\rho_B} - 1 \right|
\end{equation}
to assess dynamical similarity. By construction, \(\mathcal{E} \in [0, 1]\), with \(\mathcal{E} = 0\) and \(\mathcal{E} = 1\) indicating high and low dynamical similarity between \(u(t)\) and \(v(t)\), respectively.

For the causal flavor of DDA, \textbf{CD-DDA} (Cross-Dynamical DDA) (see \cite{lainscsek19-101103} for more details), we compare the model error of one time series to the model error of a combined model. To be more precise, if we want to test if $v(t)$ has a causal influence on $u(t)$, we compute the model errors 
\begin{equation}
\begin{array}{lll}
& {\bf \dot{u}} = {\bf M_u} \, {\bf A} & \Rightarrow \rho_u \\
\text{and} & & \\
& {\bf \dot{u}} = ({\bf M_u},{\bf M_v}) \, {\bf D}  & \Rightarrow \rho_{uv} 
\end{array}
\label{eq9}
\end{equation}
where we stack the delay matrices horizontally in the second equation. 
Note, that for time series that were generated by different mechanisms, the DDA models (delays and/or model forms) and therefore the delay matrices ${\bf M_u}$ and ${\bf M_v}$, might be different.
We then obtain the difference of the errors
\begin{equation}
\mathcal{C}_{uv} = | \rho_u - \rho_{uv} |
\label{eq10}
\end{equation}
to test for causality $\mathcal{C}_{uv}$ from $v(t) \rightarrow u(t)$. If $v(t)$ is not causally connected to $u(t)$, the error will not change since the additional terms in the model are irrelevant, but will reduce the error if a causal connection $v(t) \rightarrow u(t)$ exists.

\subsection{CDR}

There are problems with the above approach: (i) If the two time series $u(t)$ and $v(t)$ are dynamically similar, but not causally connected a false positive in $\mathcal{C}$ could be detected. In \cite{lainscsek21-103108} we solved this problem by  multiplying $\mathcal{C}$ with $\mathcal{E}$. As we noticed later, this only solves the problem if there are other causal connections in the network. For an unconnected network of independent time series, false positives can arise and are difficult to identify. CDR can detect those. (ii) Another problem is synchronization. If two time series are synchronized the direction of causality cannot be determined, since both, $\mathcal{C}_{uv}$ as well as $\mathcal{C}_{vu}$, increase even if there is a unidirectional causal connection. CDR can identify such connections.
\begin{wrapfigure}{R}{0.4\textwidth}
    \centering
    \includegraphics[width=0.4\columnwidth]{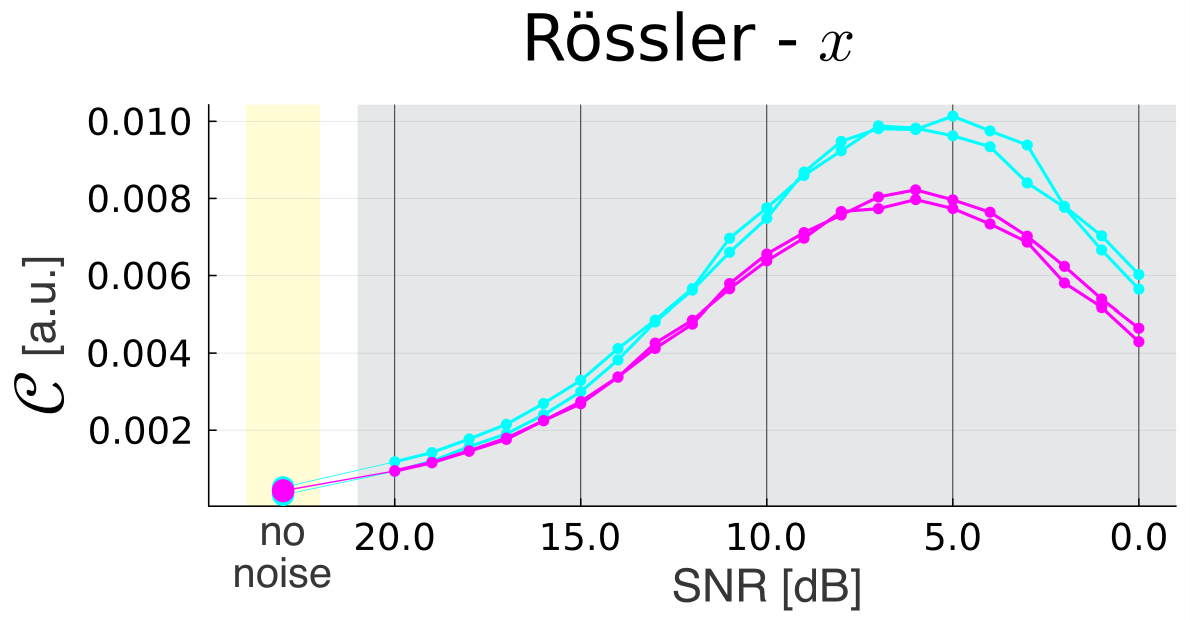}
    \\
    \includegraphics[width=0.4\columnwidth]{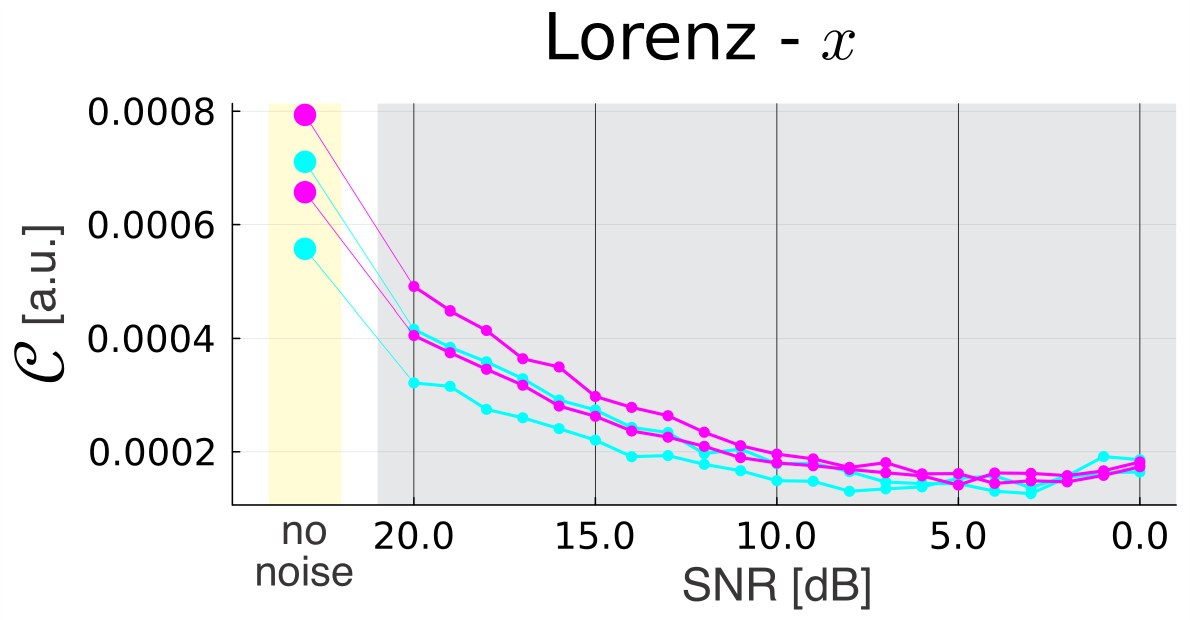} 
    \caption{Causal dynamic recurrence test to identify data where false positives can be detected with CDR. We compute $\mathcal{C}$ for independent data windows between two similar time series (magenta) and between independent data windows of the time series itself (cyan). Each dot shows the mean for all data window combinations.}
    \label{fig:CD_test}
\end{wrapfigure}

CDR is not a new flavor of DDA, but a mechanism to identify false connections by adding Gaussian white noise to data. Such false positives only appear if the data are similar, i.e.\ coming from a similar dynamical process.

To determine whether all data exhibit CDR we use the \textbf{causal dynamic recurrence test}:
Time series data might have similar sequences over time. For some dynamical systems, such sequences occur more frequently than others. For example for the $x$-timeseries of the R\"ossler system we expect more such repetitions than for the $x$-timeseries of the Lorenz system. We therefore compute $\mathcal{C}$ between five consecutive data windows from the same time series. There is no causality, but if the segments are similar to each other, 
we expect false positives. Furthermore, adding noise to such false positive cases
 can make the delay matrices ${\bf M_u}$ and ${\bf M_v}$ in Eq.~[\ref{eq9}] dynamically more similar 
and therefore ${\bf M_v}$ has an increased effect on the error $\rho_{uv}$, causing $\mathcal{C}$ in Eq.~[\ref{eq10}] to increase. If too much noise is injected, $\mathcal{C}$ will decrease again.
For the Lorenz $x$-timeseries temporally disconnected data windows or independent $x$-time series are rarely similar to each other and therefore added noise has no effect, while for the $x$-timeseries of the R\"ossler system there should be an effect.

To demonstrate this we ran an experiment using $x$-timeseries of the Lorenz and R\"ossler systems. The equations and parameters of the two systems are listed in the appendix.
We generated two time series for each system and computed $\mathcal{C}$ between five overlapping (shifted by 100 data points) consecutive data windows of 3000 data points each (sliding the middle window across the time series) from the same time series and between the two dynamically similar time series from each system.
There is of course no causality between such data windows.
We then added Gaussian white noise of varying intensity to the time series data (the ``signal'' in what follows) leading to a formal definition of the ``Noise SNR'':
\begin{equation} \label{eq:SNR}
\text{SNR}_{\text{dB}} = 10 \log_{10} \ffrac{\text{var}(\text{signal})}{\text{var}(\text{noise})}.
\end{equation}
We set \(\text{SNR}_{\text{dB}} = [20\text{dB}, 19\text{dB}, \dots, 1\text{dB}, 0\text{dB}]\).
Fig.~\ref{fig:CD_test} shows the results: The cyan curves (both timeseries) correspond to causality within a time series and the magenta curves are between random independent data segments of the two time series (two directions). The cyan and magenta curves have similar values.
In the experiments using the R\"ossler system, the CDR effect is present while using the Lorenz system no such CDR response is observed. Additionally, the magnitude of $\mathcal{C}$ via Lorenz data is much lower than via R\"ossler data.
This means we can add noise and test for false connections when the dataset is derived from a Rossler system. For cases, where there is no CDR, false positives do not exist. 
\begin{figure}[hbt]
    \centering
   \includegraphics[width=0.8\columnwidth]{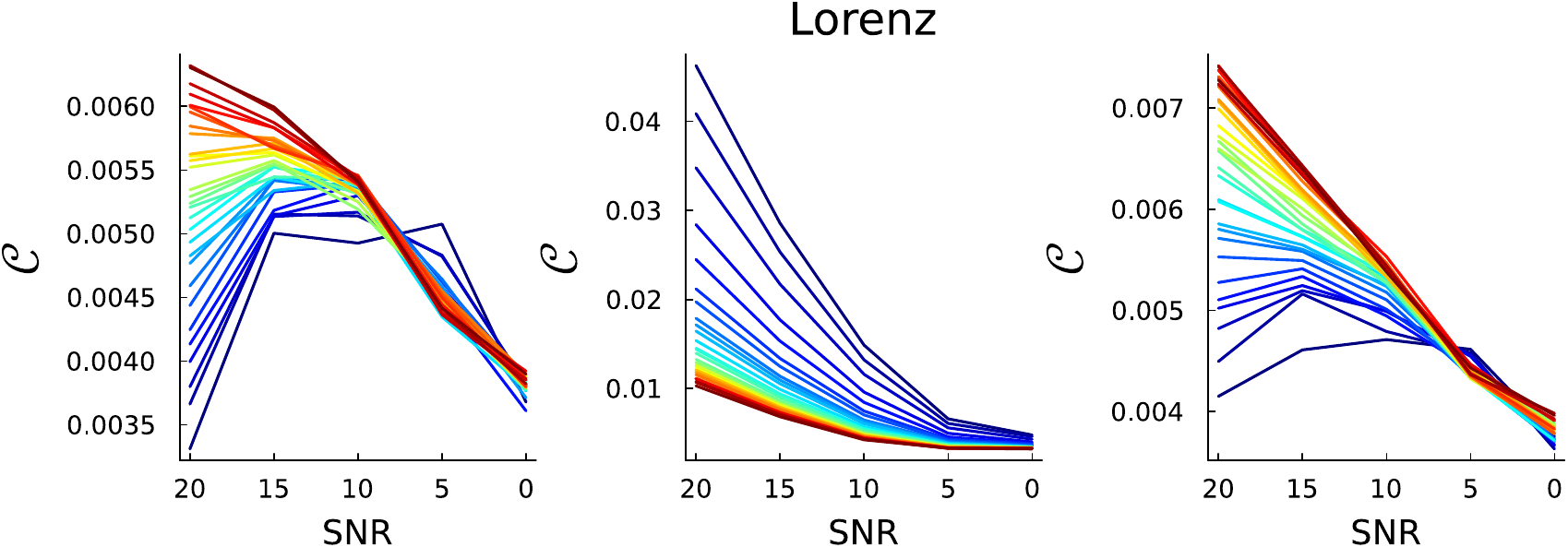}
    \caption{Dependence of CDR on self-similarity: We generated a time series from the $x$-component of the Lorenz system with 15000 data points (see appendix). One pseudo period as on average around 170~$\delta t$. We therefore use a window length of 170~$\delta t$ and a window shift of 1~$\delta t$ and compute $\mathcal{C}$ as well as the delay matrix $({\bf M_u},{\bf M_v})$ in Eq.~[\ref{eq9}]
    between all data windows resulting in around 219 million combinations. We then sort $\mathcal{C}$ for all combinations by the Pearson correlation between the first, second, and third columns of ${\bf M_u}$ and ${\bf M_v}$ as a measure of self-similarity between independent data windows. We then take the mean of 15000 consecutive sorted $\mathcal{C}$'s and show the resulting plots (blue to red for the first 20 of around 15000 such lines) as functions of the SNR. We get the CDR effect only for around 7\% of all $\mathcal{C}$'s for the first columns in the delay matrices and around 3\% for the last column. There is no CDR effect in the middle column. 
    }
    \label{fig:Lor_SelfSim}
\end{figure}

CDR depends on dynamical similarity between the time series under investigation. In Fig.~\ref{fig:Lor_SelfSim} we look at self-similarity between data segments within a single time series, similar to the experiment in Fig.~\ref{fig:CD_test}. In this experiment we use short time windows of only one pseudo period to see an effect. We then look at how $\mathcal{C}$ changes with similarity of the data windows assessed by the Pearson correlation between the columns in the ${\bf M_u}$ and ${\bf M_v}$ matrices. The first columns are $u_{\tau_1}$ and  $v_{\tau_1}$, the second columns are $u_{\tau_1}^2$ and  $v_{\tau_1}^2$, and the third columns are $u_{\tau_1}^2 u_{\tau_2}$ and  $v_{\tau_1}^2 v_{\tau_2}$. For the first and third columns we see the CDR effect for only a small percentage of data window combinations. For longer data windows this effect gets less and less frequent as less data windows will be similar to each other.
In the R\"ossler system we have no symmetry and data windows will be self similar in a high percentage and therefore we get CDR for dynamically similar time series.

\section{Possible Connections to Stochastic Resonance}

At this point we offer a digression to compare CDR and another well-studied noise-mediated effect in nonlinear dynamics. At first glance, CDR might appear to be a manifestation of the much-studied ``Stochastic Resonance'' (SR) effect \cite{bulsara96-39}, because of the Causality vs. Noise profile, as seen in Fig.~\ref{fig:SR_explain}. However, SR is a nuanced, often quite ``delicate'' phenomenon, prone to mis-labelling; hence it is instructive to provide a brief overview of the effect, with emphasis on variations that might be germane to the system at hand.

SR, in one of its fundamental manifestations, occurs in a deterministically driven nonlinear dynamical system with an intrinsic potential energy barrier (predicated by the system dynamics) that the state point must cross for an applied signal to be detected. This process is aided by noise when the signal amplitude is below the deterministic crossing threshold i.e., the threshold above which the signal can generate crossing events in the absence of noise. There is, also, a noise-mediated hopping rate over the barrier, in the absence of the signal. For the basic case of a tone input signal, a matching between these deterministic and stochastic rates,leads to a definition of SR in line with the commonly accepted physics definition of a ``resonance''. 

The above characterization of SR is strictly applicable to time-periodic input signals or combinations thereof, and comes with an important limitation. The frequency matching condition is only approximately satisfied when using a spectral response function, e.g. the output SNR measured at the input signal frequency, to characterize the system. For this case one realizes the ``resonance'' (the maximum of the output SNR as a function of the input noise) when the noise intensity approximately matches the energy barrier height; this resonance behavior is, however, only weakly dependent on the signal frequency, and does not rise to the above-mentioned physics definition of a resonance. Frequency matching resonance is more precisely characterized by considering  an alternative (and more basic) quantification of the system response in terms of the level crossing statistics of the state point \cite{gammaitoni95-1052}; a good overview is provided in \cite{bulsara96-39}. 

There exist more generalized cases of a noise-induced maximum in the response  of a nonlinear system to non-sinusoidal (e.g. wideband or aperiodic) input signals. The response measure is, typically, a correlation between the input signal and a (noise-dependent) function of the output, e.g.\ an information transfer function \cite{bulsara96-R2185, stemmler96-687}, a Kulbach Entropy \cite{anishchenko99-7} or some other, suitably defined, measure. A maximum in the Response Function vs. Noise profile typically arises via a noise-mediated  property of the underlying system. As an example, consider the threshold system (leaky integrate-fire model) of ref \cite{bulsara96-R2185} subject to a wideband signal. Here, the competing effects of low threshold crossing or ``spike'' rate (at low noise) but with greater information per spike, and a higher spike rate at higher noise but with less information per spike, lead to a ``crossover'', characterized by a maximum in the information throughput at a critical noise intensity, between these regimes. Quantitative features (e.g. the slopes of the ascending and descending segments of the response curve, and the location of the maximum) are, of course, dependent on the choice of system parameters. 

For the system at hand, we note that the data set used to generate the response of Fig.~\ref{fig:SR_explain} was generated via a R\"ossler system, with Gaussian white noise superimposed on it. So the input ``signal'' is a noisy chaotic signal, demanding more generalized response measures to study its implications for the system dynamics. We note that, for a single R\"ossler element, a mechanism for generating SR-like behavior has been studied \cite{neiman07-1442, gingl95-191}  using the intermittency generated by deviations from the phase-locked state. The SR-like behavior is characterized by an enhancement, via the internal chaos, of a suitably chosen measure of the response at an intrinsic frequency of the underlying system; this effect can be enhanced by small amounts of noise. Importantly, no external time-periodic signal need be added to the system, in contrast with conventional SR. Additionally, an energy barrier (as described above) is not necessary, with the ``resonance'' corresponding to a synchrony between time-scales associated with the (weak) noise-generated behavior and the internal oscillatory behavior in the underlying system in the absence of noise. This behavior is often classified under the broad rubric of ``Coherence Resonance'' (CR) in the physics repertoire (see \cite{neiman07-1442, gingl95-191} for an overview). 

The above considerations suggest an interesting connection between Coherence Resonance and CDR in a system which is ``noisy-chaotic''. Additionally, when the input data is generated via coupled chaotic systems, the coupling parameter(s) can generate a driver-response relationship in the inter-element dynamics, leading to an enhancement (or de-enhancement, depending on the system parameters) of the SR-like behavior. It is tempting to speculate that the above considerations might explain the behavior of the curves in Fig.~\ref{fig:SR_explain}, specifically the monotonic increase of the Causality at low noise. Under this hypothesis, CDR could be considered a ``non-dynamic'' \cite{gingl95-191} resonance in a noisy chaotic system (the data set derived from the R\"ossler system and laced with Gaussian white noise). The ``resonance'' would not require a physical ``energy barrier'' threshold and it would not require the system to be driven by a tonal signal. 


\section{Two unidirectionally coupled R\"ossler systems}

To highlight the problem of the connection of causality and synchrony we revisit and expand the example of diffusive coupling between two R\"ossler systems \cite{roessler76-397} as used in Palu\v{s} and Vejmelka~\cite{palus2007}. 
\begin{equation}\label{eq:palus}
\begin{array}{ll}
\boldsymbol{D} &
\left\{ 
\begin{array}{rcl}
\dot{x}_{1} & = & - \omega_1 y_{1} - z_{1} \\
\dot{y}_{1} & = & \omega_1 x_{1} + a y_{1} \\
\dot{z}_{1} & = & b_0 - c z_{1} + x_{1} z_{1} \\ [1ex]
\end{array} 
\right. \\
\boldsymbol{R} &
\left\{ 
\begin{array}{rcl}
\dot{x}_{2} & = & - \omega_2 y_{2} - z_{2}  + \epsilon (x_{1} - x_{2})
\\
\dot{y}_{2} & = & \omega_2 x_{2,j} + a y_{2} \\
\dot{z}_{2} & = & b - c z_{2} + x_{2} z_{2} \\ 
\end{array}
\right.
\end{array}
\end{equation}
where $b=0.2$, $c=10$,  $\omega_1=1.015$, $\omega_2=0.985$, $a_1 = a_2 = 0.15$, and $\epsilon$ is the coupling strength.
$\boldsymbol{D}$ is the driver and $\boldsymbol{R}$ is the response system.

\subsection{Response system}
\begin{wrapfigure}{R}{0.42\textwidth}
    \centering
    \includegraphics[width=0.4\columnwidth]{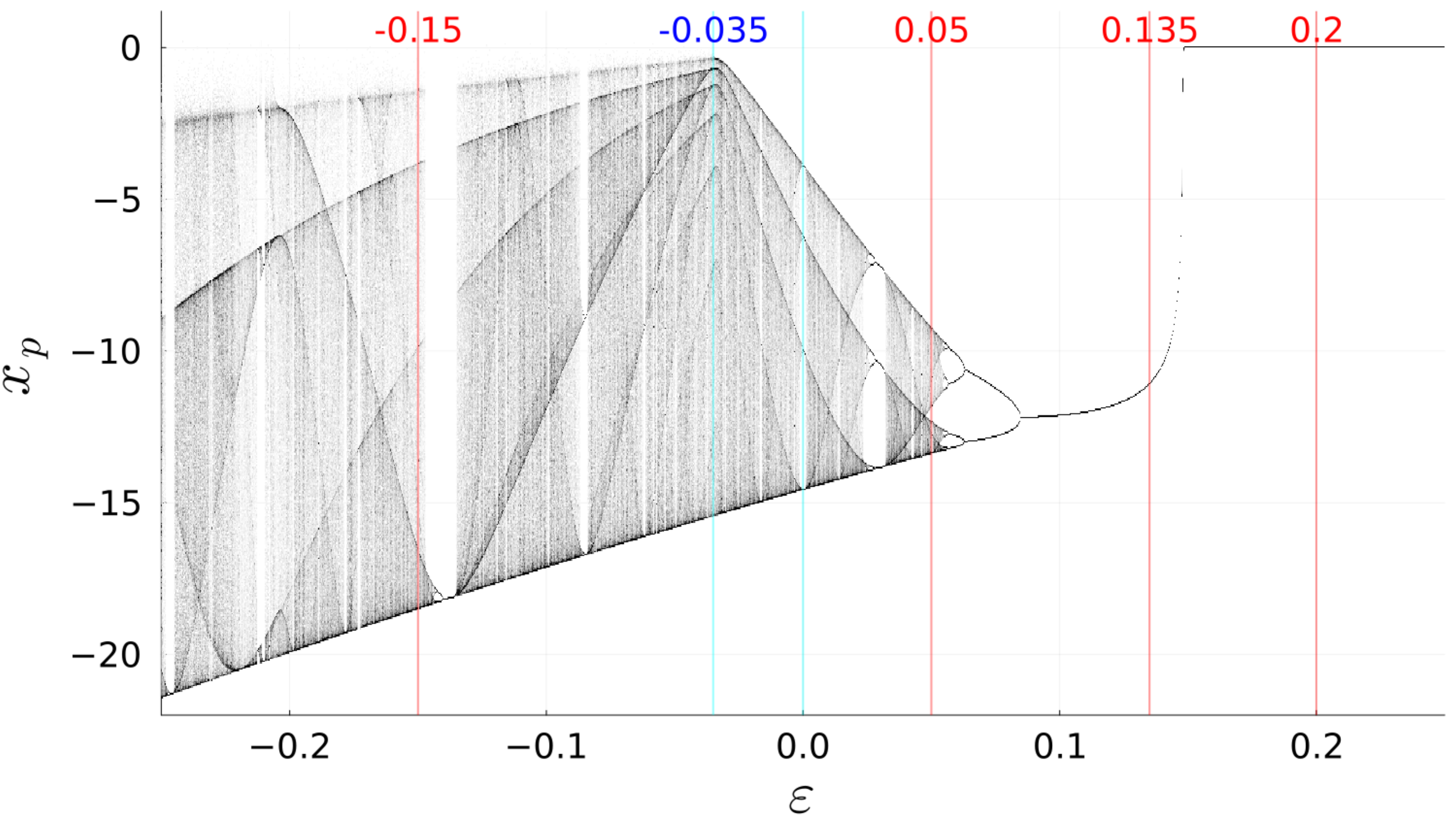}
    \caption{Bifurcation diagram of the modified response system [\ref{eq:palusR}] without input from the driver system ($x_1 = 0$).}
    \label{fig:PalusR}
\end{wrapfigure}
In \cite{lainscsek11-046205} Lainscsek showed that the R\"ossler system (above response system with $\epsilon=0$) has three effective bifurcation parameters using the Ansatz Library approach. All terms of the system except $a$ and $c$ are connected. We repeat this analysis for the response system and get
\begin{equation}\label{eq:palusR}
\begin{array}{ll}
\boldsymbol{R} &
\left\{ 
\begin{array}{rcl}
\dot{x}_{2} & = & \boxed{-\epsilon x_{2}} - n \omega_2 y_{2} - k m z_{2}  + m  \, \boxed{\epsilon x_{1} }
\\ [1ex]
\dot{y}_{2} & = & \ffrac{1}{n} \omega_2 x_{2,j} + a y_{2} \\  [1ex]
\dot{z}_{2} & = & \ffrac{1}{k} b - c z_{2} + \ffrac{1}{m} x_{2} z_{2} \\ 
\end{array}
\right.
\end{array}
\end{equation}
where we additionally split up the two diffusive coupling terms (in boxes).
The additional term $-\epsilon x_{2}$ in the first equation is an additional bifurcation parameter, while the coupling term $\epsilon x_{1}$ is connected to all the other parameters of the system. A more detailed explanation can be found in the appendix. Note, that the parameters $m$, $n$, and $k$ are only added to show the connections between the parameters and will be here set to 1.

As a first step we look at this system without coupling from the driver system ($x_1 = 0$) and plot the bifurcation diagram as function of $\epsilon$ in Fig.~\ref{fig:PalusR}.
The additional vertical lines will be explained in the DDA part of this section below.
The attractor is funnel chaotic for around $\epsilon < -0.035$, then it is spiral chaotic for $-0.035 < \epsilon < 0.055$, as the driver system.
For $\epsilon > 0.055$ the system is periodic and for $\epsilon > 0.149$ we get a point attractor.

We therefore could expect that for $\epsilon < -0.035$ complete synchronization is harder to achieve because driver and response system are more different than for 
$\epsilon > -0.035$. The results in the next section confirm this but a more thorough investigation is part of our ongoing research and would go beyond the scope of this paper.

\subsection{DDA analysis of the coupled system}

The DDA results were previously presented in our CD-DDA paper \cite{lainscsek19-101103}. Later we by mistake flipped the two coupling terms to $x_{2} - x_{1}$ in above equation and realized that there is synchronization for $x_{1} - x_{2}$, but not for most of $x_{2} - x_{1}$. We therefore ran the experiment in \cite{lainscsek19-101103} again for $-0.25 \le \epsilon \le 0.25$ to show both cases in one simulation. As synchrony measure we use the Kuramoto order parameter \cite{kuramoto02-380}
$\mathcal{O} = \frac{1}{N} \sum_{j=1}^N e^{i \theta_j}$
where $N=2$ is the number of oscillators in [\ref{eq:palus}].

\begin{figure}[hbt]
    \centering
    \includegraphics[width=0.4\columnwidth]{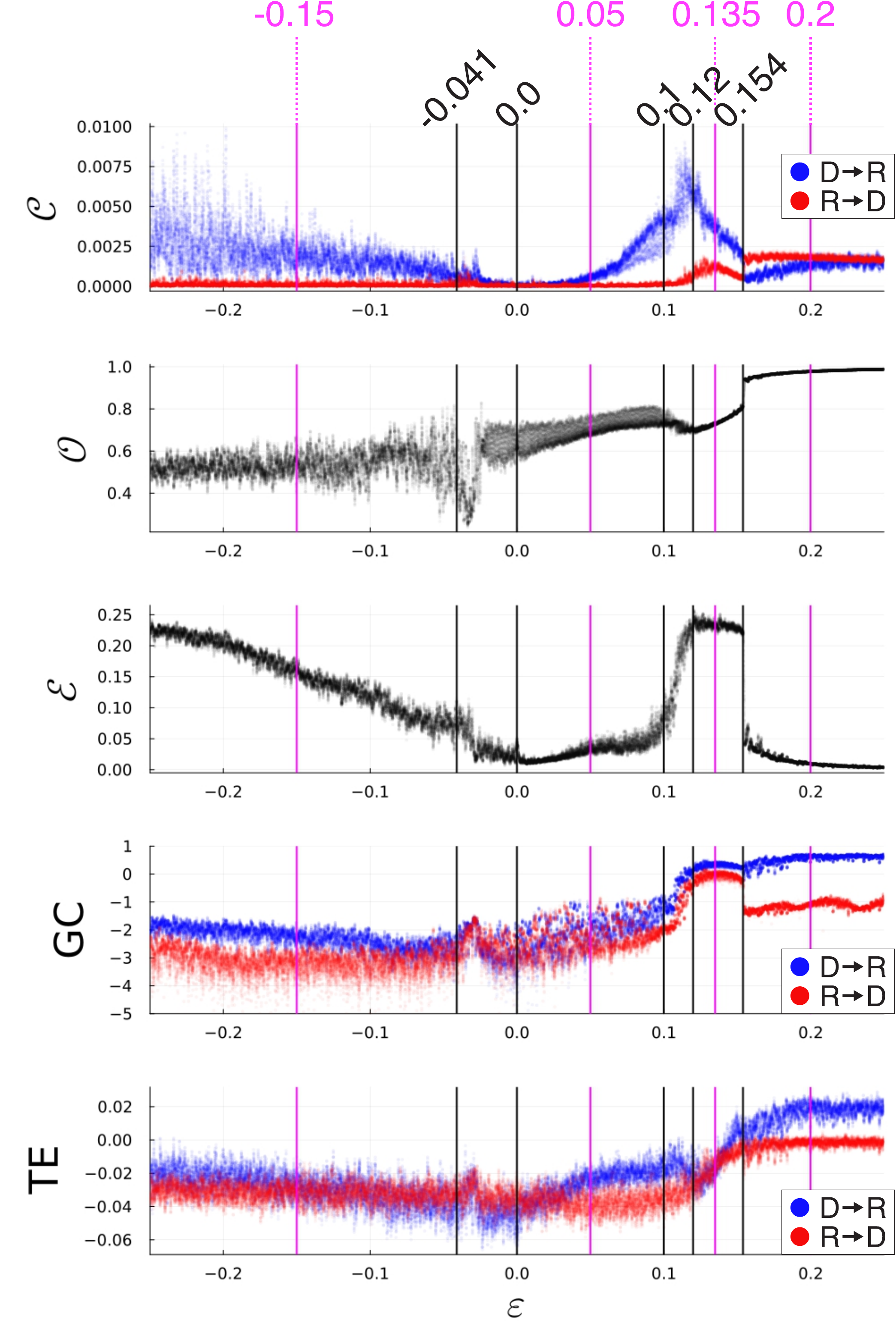}
    \caption{DDA causality $\mathcal{C}$, Kuramoto order parameter $\mathcal{O}$, dynamical similarity $\mathcal{E}$ Granger causality (GC), and transfer entropy(TE) for the two unidirectionally coupled R\"ossler systems [\ref{eq:palus}] as functions of the coupling strength $\epsilon$. GC is shown on a logarithmic scale on the $y$-axis due to large variations in scale.  The black vertical lines at (-0.041, 0.12, 0, 0.154) indicate changes in behavior and CDR for values marked with magenta vertical lines at (-0.15, 0.05, 0.135,0.2) is shown in Fig.~\ref{fig:Palus_CDR}.}
    \label{fig:Palus_COE}
\end{figure}
Additionally, we compare our results with GC  and TE.
For GC we use the Bayes Information Criterion to determine the auto-regressive order and we use the \texttt{armorf.m} function from the BSMART toolbox for AR parameter estimation \cite{bsmart2008}. For TE we use MATLAB TRENTOOL (TRansfer ENtropy TOOLbox) \cite{trent2011}, using the Ragwitz criterion to select the optimal embedding dimension and time delay. 

\begin{wrapfigure}{R}{0.42\textwidth}
\vspace*{5mm}
    \centering
    \includegraphics[width=0.4\columnwidth]{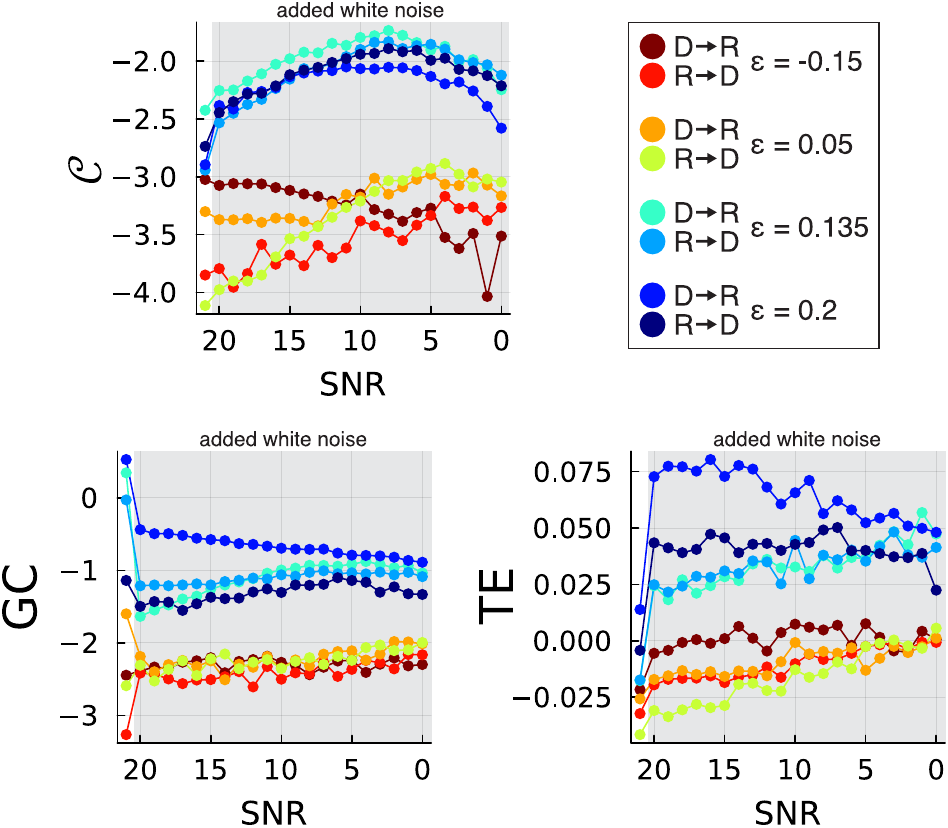}
    \caption{Causality $\mathcal{C}$, GC, and TE for no noise (left dots) and for added white noise with a SNR between 20 and 0~dB (gray boxes). The $y$-values for $\mathcal{C}$ and GC are on a logarithmic scale.}
    \label{fig:Palus_CDR}
\end{wrapfigure}

The two R\"ossler systems [\ref{eq:palus}] were integrated with a step size of 0.05 and down-sampled by a factor of two to be consistent with~\cite{lainscsek19-101103,lainscsek23-123136}. 
The initial conditions for each time series are chosen randomly.
While integrating the system, we incrementally increase the coupling parameter 
$\Delta \epsilon = 1\text{e}^{-7}$ 
at every integration step between -0.25 and 0.25 instead of multiple time series for each $\epsilon$ as used in 
\cite{lainscsek19-101103,lainscsek23-123136}. The results for $0 < \epsilon \le 0.25$ are very similar but computations are simpler and faster.
This results in 5,000,000 data points for each of the two time series of the $x$-variables in [\ref{eq:palus}]. 

In Fig.~\ref{fig:Palus_COE} we compare DDA causality $\mathcal{C}$, the Kuramoto order parameter $\mathcal{O}$, dynamical similarity $\mathcal{E}$, GC (shown on logarithmic scale to account for huge changes of the values), and TE as functions of the coupling strength $\epsilon$.
$\mathcal{O}$ increases non-monotonically as $\epsilon$ increases. 
For $-0.02 \le \epsilon \le 0.1$ the two time series $x_1$ and $x_2$ from [\ref{eq:palus}] slip in and out of phase synchronization. For this $\epsilon$-range dynamical ergodicity is low indicating dynamically similar $x_1$ and $x_2$. 
$\mathcal{C}$ increases correctly, GC flip-flops, and TE increases correctly for  $0 \le \epsilon \le 0.1$.
$\epsilon = 0.12$ indicating the onset of generalized synchronization (see \cite{palus2007} and \cite{lainscsek19-101103}).
For $0.12 \le \epsilon \le 0.154$ $x_1$ and $x_2$ show phase synchronization, but not amplitude synchronization as indicated by a high $\mathcal{O}$ and a high $\mathcal{E}$. $\mathcal{C}$, GC, and TE all indicate causality in the correct direction even in the presence of phase synchronization. For  $0.154 \le \epsilon \le 0.25$ $x_1$ and $x_2$ are synchronized and dynamically similar as indicated by a high $\mathcal{O}$ and a low $\mathcal{E}$ (high dynamical similarity). 
For $-0.25 \le \epsilon \le -0.041$ there is no phase synchronization and dynamical similarity is low. 

As mentioned above, detecting the direction of causality is a problem for dynamically similar or synchronized time series resulting in false positives. We therefore add Gaussian white noise 
with SNR values between 20 and 0~dB in steps of 1~dB  
to the data for the example $\epsilon$-values marked in magenta in Fig.~\ref{fig:Palus_COE} and show the results in Fig.~\ref{fig:Palus_CDR}. All coupling strengths except $-0.25 \le \epsilon \le -0.041$ result in some level of synchronization as discussed above and therefore might show false positives. Only for $-0.25 \le \epsilon \le -0.041$ and $x_1 \rightarrow x_2$ causality is not obscured by similarity or synchrony. 
In Fig.~\ref{fig:Palus_CDR} for $\mathcal{C}$ only the dark red line for $-0.25 \le \epsilon \le -0.041$ and $x_1 \rightarrow x_2$ goes down as noise increases. All other $\epsilon$-values show some CDR which is the strongest (lines in different shades of blue) above the onset of general synchronization at around $\epsilon = 0.12$.
This effect cannot be seen for GC or TE.
For $\epsilon > 0.154$ the Kuramoto order parameter is nearly one and we have nearly perfect phase synchronization. CD-DDA then cannot detect the direction of causality, but GC and TE can.

For the remainder of this manuscript we only consider $\epsilon = -0.15$, where we have dynamically different driver and response systems and no synchronization.

\section{Network of Coupled R\"ossler Systems}%
\label{sec:networks}

To test the effect of noise on CD-DDA, we couple seven R\"{o}ssler systems using  the following coupling
\begin{equation}\label{eq:ros}
    \begin{array}{rcl}
        \dot{x}_{n} & = & -y_n - z_n + \sum\limits_j \epsilon (x_n - x_j) \\
        \dot{y}_{n} & = & x_{n} + a_n \, y_{n} \\[1ex]
        \dot{z}_{n} & = & b_n + c_n z_{n} + x_{n} z_n 
    \end{array} 
\end{equation}
with $n=1,2,\dots,7$ and $x_j$ is the $x$-component of another system. The values for $a_n$, $b_n$, and $c_n$ are listed in
\cref{tab:ros}. There are three dynamically different groups of systems: (i) systems (1,2,3), (ii) systems (4,5,6), and (iii) system 7,
highlighted by three different colors in \cref{tab:ros}. The coupling constant $\epsilon$ is either 0 or 0.15 depending on which systems
are coupled. The seven R\"ossler systems were integrated with a step size of 0.05 and down-sampled by a factor of two to be consistent
with~\cite{lainscsek19-101103,lainscsek23-123136} as in the previous section. 
The initial conditions for each time series are chosen randomly.

\begin{table}[hbt]
    \centering
    \includegraphics[width=.3\textwidth]{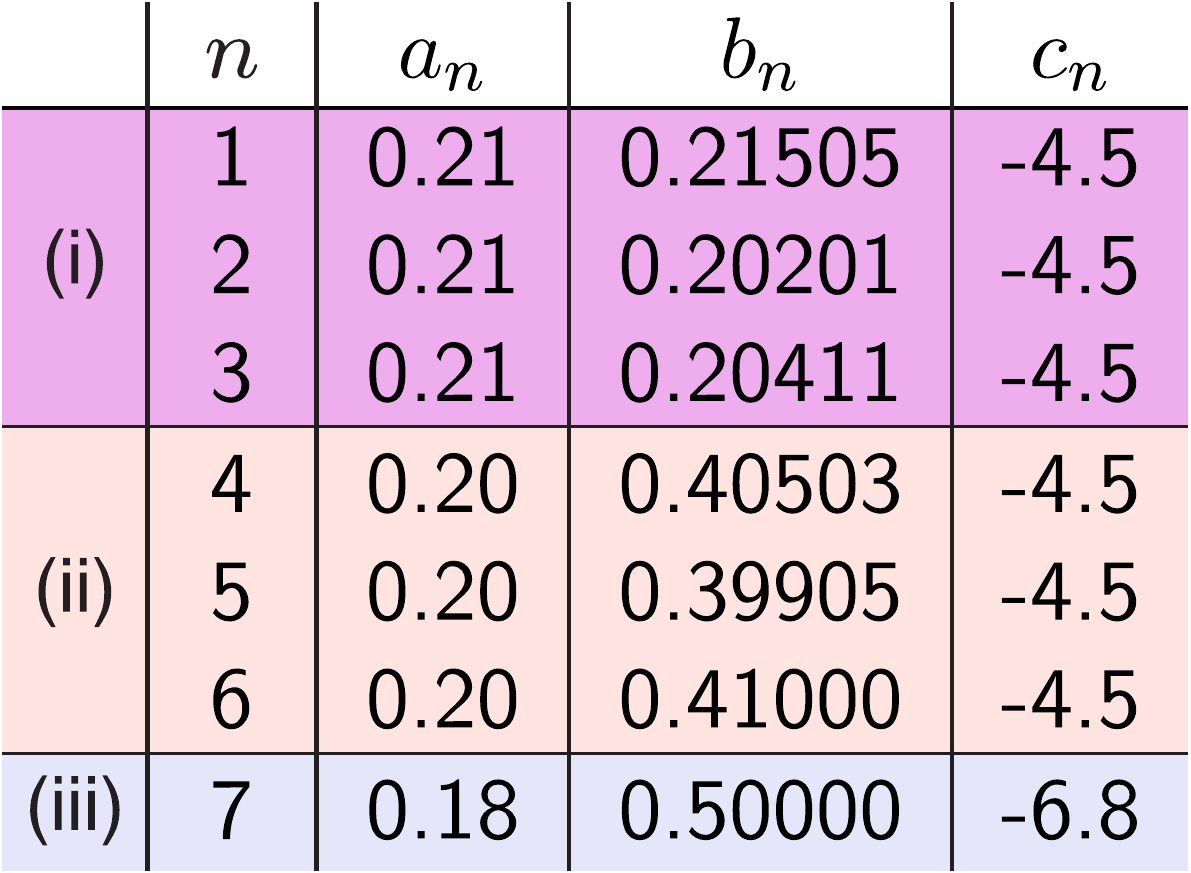}
    \caption{Parameters of the seven R\"ossler systems. {\normalfont The three colors indicate the three dynamical regimes (i), (ii), and (iii).}}
    \label{tab:ros}
\end{table}

\subsection{First network example}%

\begin{figure}[hbp]
    \centering
    \includegraphics[width=0.4\columnwidth]{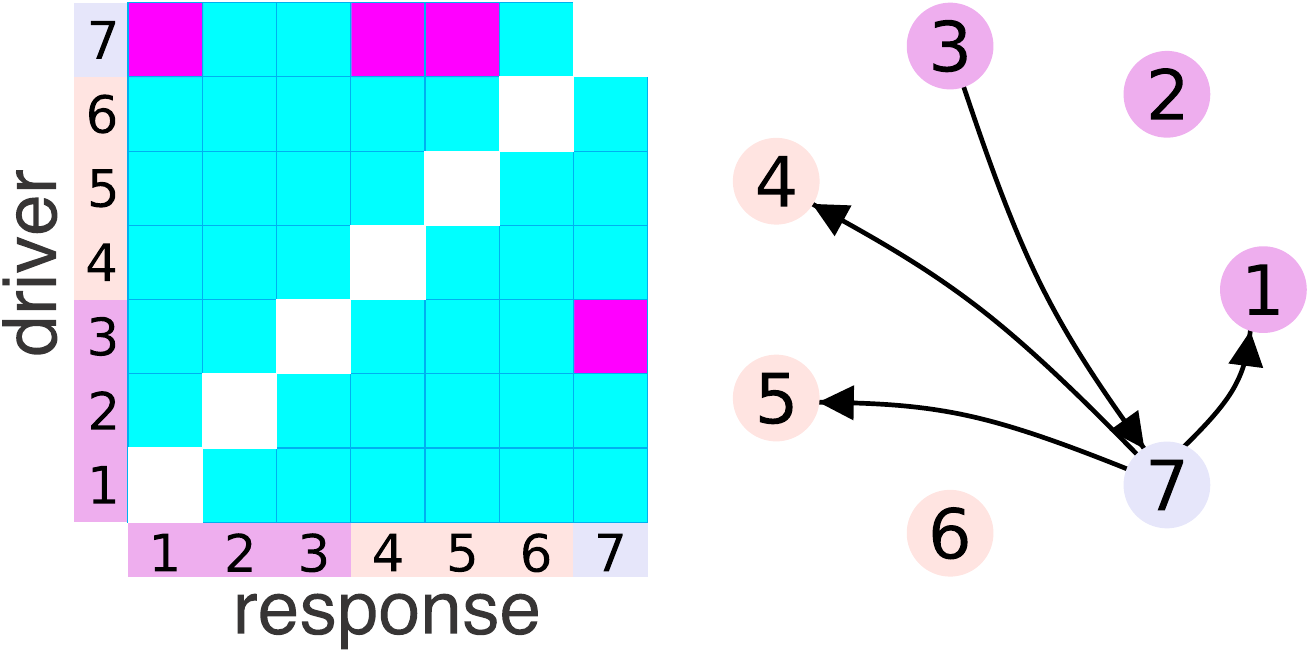}
    \caption{Adjacency matrix and network motif of ground truth. The connections are indicated as magenta boxes.}
    \label{fig:Motifs_truth_1}
\end{figure}

For the first example the network of the seven R\"ossler systems
(R1,R2,...,R7)
in \eqref{eq:ros} reads as
\begin{equation}    
\begin{array}{rl}
\rotatebox[origin=c]{90}{(i)} &
\left\{
\begin{array}{l}
\CR{ + 0.15 (x_1 - x_7)}{1} \\
\CR{}{2} \\
\CR{}{3} 
\end{array}
\right.
\\ \\
\rotatebox[origin=c]{90}{(ii)} &
\left\{
\begin{array}{l}
\CR{ + 0.15 (x_4 - x_7)}{4} \\
\CR{ + 0.15 (x_5 - x_7)}{5} \\
\CR{}{6} 
\end{array}
\right.
\\ \\
\rotatebox[origin=c]{90}{(iii)}  &
\left\{
\begin{array}{l}
\CR{ + 0.15 (x_7 - x_3)}{7} 
\end{array}
\right.
\end{array}
\label{eq:RosEx1}
\end{equation}
with the values for $a_n$, $b_n$, and $c_n$ ($n = (1,2,\dots,7)$) as listed in \cref{tab:ros}. (i), (ii), and (iii) denote the different
dynamical regimes. The adjacency matrix and network motif of ground truth in \eqref{eq:RosEx1} is shown in \cref{fig:Motifs_truth_1}.
The three different colors in the nodes of the motif denote the three dynamical regimes (i), (ii), (iii).

We computed the DE-DDA ergodicity matrix and the CD-DDA adjacency matrix for all pairwise combinations of the $x_n$ ($n = (1,2,\dots,7)$)
time series of the seven systems from \eqref{eq:RosEx1}.
As has been discussed in~\cite{lainscsek23-123136}, there can be false positive connections caused by the dynamical similarity of the data (lower ${\cal E}$ values). While in~\cite{lainscsek23-123136} we used the multiplication ${\cal C} \times {\cal E}$ to remove such false connections, here we use a step in between to cover more cases.

\begin{figure}[htb]
    \centering
    \includegraphics[width=0.5\columnwidth]{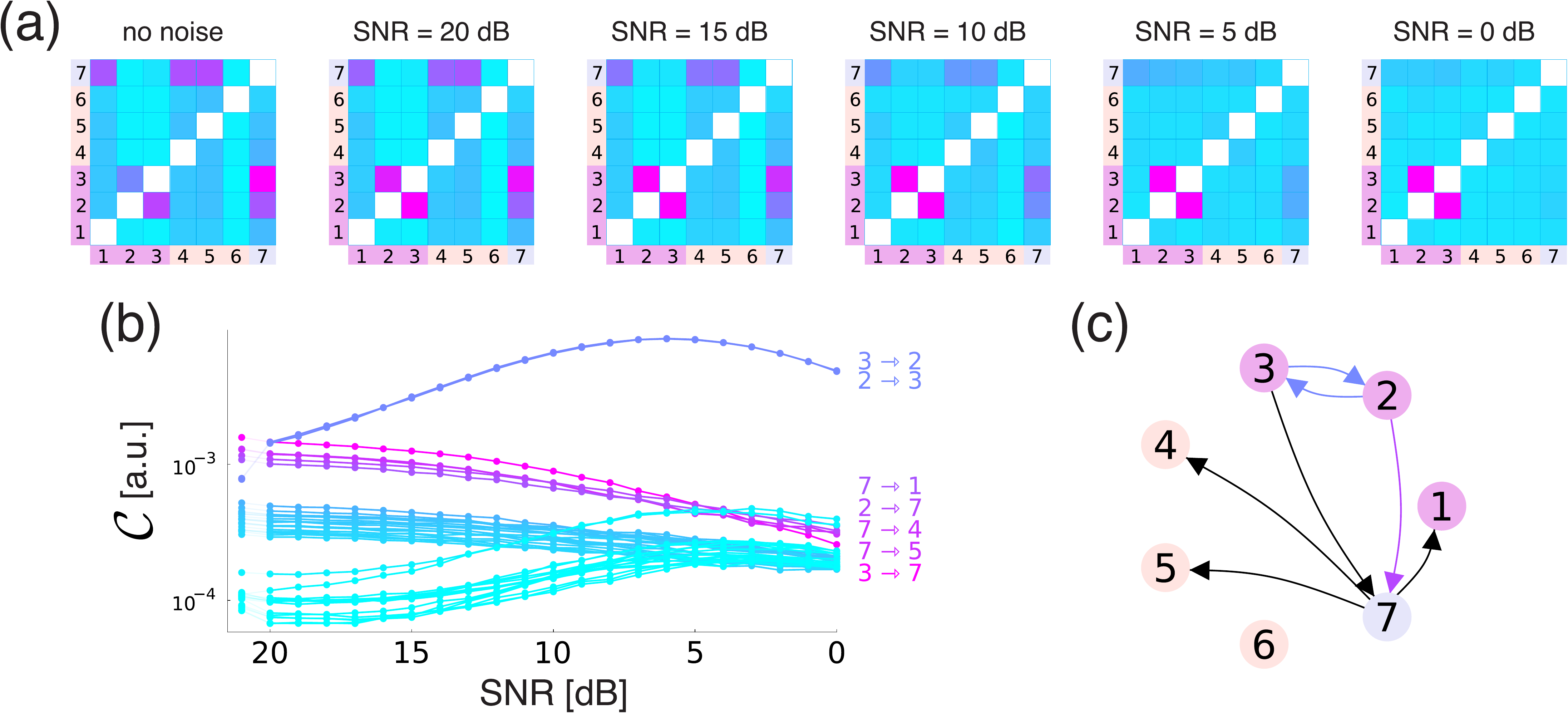}
    \caption{Numerically obtained adjacency matrices (top), dependence of ${\cal C}$ on SNR, and the numerical network motif (bottom).
    \textsf{(a)}:~The plot on the left shows the results using noise free data while the plots to the right are from data with additive white noise such that \(\text{SNR} = 20\text{dB}, 15\text{dB}, 10\text{dB}, 5\text{dB}, 0\text{dB}\).
    \textsf{(b)}:~Causality as function of added white noise. The first values on the left are for the noise free case, followed by results from noise data with \(\text{SNR} = 20\text{dB}, 15\text{dB}, 10\text{dB}, 5\text{dB}, 0\text{dB}\). The colors correspond to the colors from the noise-free adjacency matrix.
    \textsf{(c)}:~Numerically computed network motif. False positives are shown as purple and magenta edges.}
    \label{fig:Motifs_numerical_1}
\end{figure}

To do so, additive Gaussian white noise with a signal-to-noise ratio (SNR)
between 20 and 0~dB in steps of 1~dB is injected into
the data and the adjacency and dynamical ergodicity matrices are recomputed. 
\Cref{fig:Motifs_numerical_1}(a) shows the numerically obtained adjacency matrices from data without and with added white noise. The causality values of some of the connections increase with noise. In \cref{fig:Motifs_numerical_1}(b) we show causality as function of added white noise to the data. The colors of the lines correspond to those of the noise free case in \cref{fig:Motifs_numerical_1}(a).  
For two edges of the network, $3 \rightarrow 7$ and $3 \rightarrow 2$ show very different behavior: For  $3 \rightarrow 7$ causality decreases with added noise, whereas for  $3 \rightarrow 2$  the causality initially  increases with noise and then decreases again, consistent with CDR.
The numerical network motif is shown in \cref{fig:Motifs_numerical_1}(c). The black arrows are the true connections and the purple and magenta arrows are the false positives.

\begin{figure}[htb]
    \centering
    \subfigure[Effect of additive white noise for two nodes.]{\includegraphics[width=0.4\columnwidth]{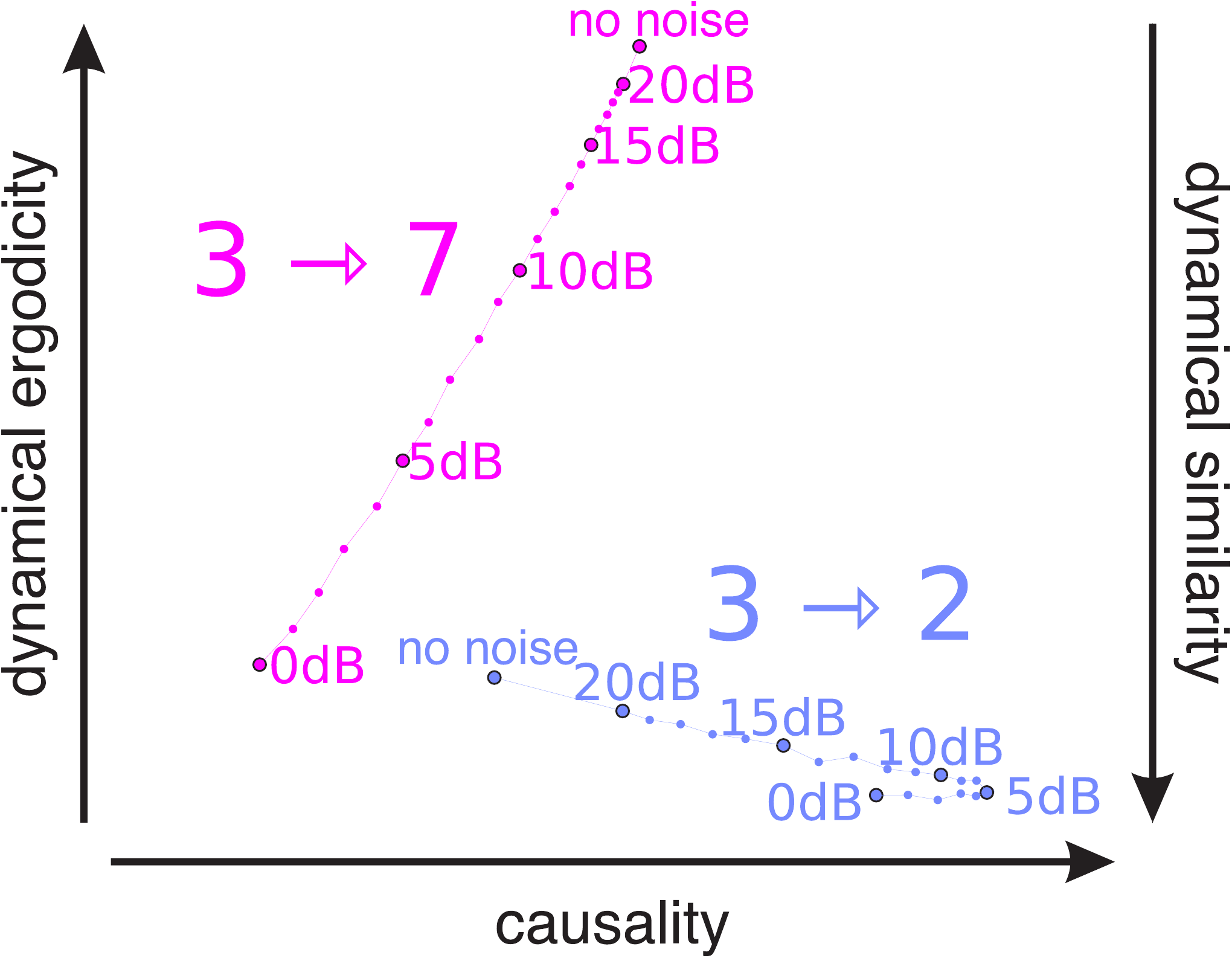}\label{fig:C_E__SR__corrcoef__1_left}}
    \\
    \subfigure[Effect of additive white noise for all nodes.]{\includegraphics[width=0.4\columnwidth]{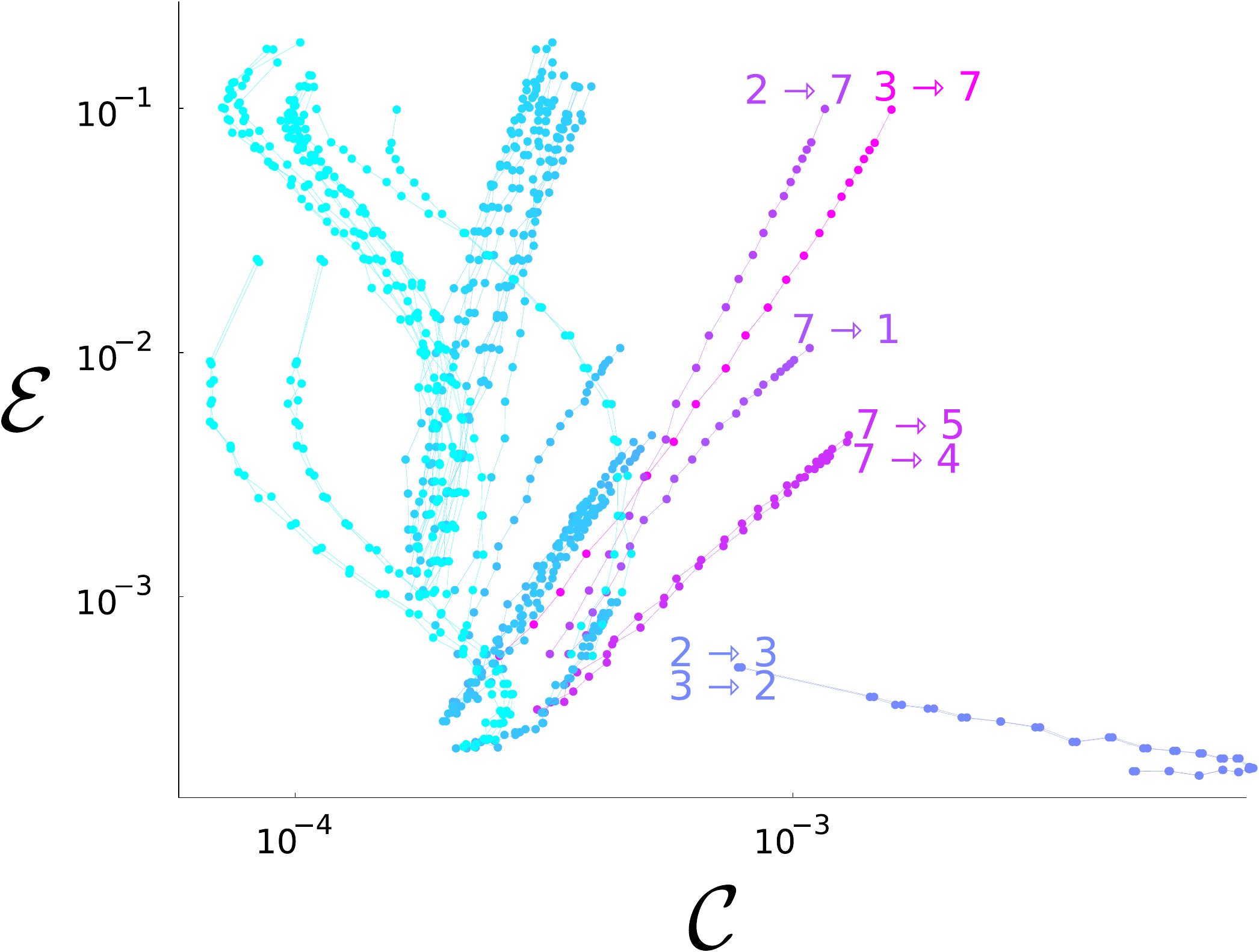}\label{fig:C_E__SR__corrcoef__1_right}}%
    \caption{Dynamical ergodicity ${\cal E}$ over causality ${\cal C}$. The colors correspond to the ones in the numerically obtained
    adjacency matrix in \cref{fig:Motifs_numerical_1}(a). On the top plot the connections $3 \rightarrow 7$ and $3 \rightarrow 2$ are shown where $3 \rightarrow 2$ is a false positive. White noise was added to the data. For $3 \rightarrow 7$ causality and dynamical ergodicity decrease with decreasing SNR. For the false positive connection $3 \rightarrow 2$ dynamical ergodicity decreases but causality increases and then decreases and we see CDR. The bottom plot shows ${\cal E}$ over ${\cal C}$ for all nodes in the network with varying SNR.}
    \label{fig:C_E__SR__corrcoef__1}
\end{figure}

In \cref{fig:C_E__SR__corrcoef__1_left} we show the effect of the added noise for two nodes of the network, namely $3 \rightarrow 7$
and $3 \rightarrow 2$, where we plot $\mathcal{E}$ over $\mathcal{C}$. The first is a correctly identified connection, while the second is a
false positive. The two connections react completely different to added white noise as can be seen in \cref{fig:C_E__SR__corrcoef__1_left}. The false positive $3 \rightarrow 2$ shows CDR: causality initially increases and then decreases as the magnitude of
added white noise increases. For the correctly identified connection $3 \rightarrow 7$ both, causality as well as dynamical ergodicity are decreasing.
In \cref{fig:C_E__SR__corrcoef__1_right} the effect of added white noise is shown for all nodes. The colors for all network nodes
correspond to the magnitude of numerically estimated causality in the noise free case and are the same as in the adjacency matrix in
\cref{fig:Motifs_numerical_1}(a) (left).

\subsubsection{Dependent Systems and CD-DDA}

There are three numerically obtained false positives where $\epsilon = 0$. The first two are $3 \rightarrow 2$ and $2 \rightarrow 3$. Both show CDR and can therefore easily be identified. 
The third false positive is $2 \rightarrow 7$. Nodes 2 and 3 are in a very similar dynamical state and therefore  
$2 \rightarrow 7$ ($\epsilon = 0$) is similar to $3 \rightarrow 7$ ($\epsilon \not= 0$), but $\mathcal{C}$ is lower.
The false positives can be explained in the following way:

For CD-DDA we consider two dynamical systems $U$ and $V$ resulting in the time series $u(t)$ and $v(t)$. In our case $U$ is the R\"ossler system R2 and $V$  is the R\"ossler system R3 in \eqref{eq:RosEx1}. The resulting time series are $u(t) = x_2(t)$ and $v(t) = x_3(t)$. As mentioned in \cref{flavors} the first step is to compute a set of features $\mathbf{A}_u = (a_1, a_2, a_3)_u$ with 
\begin{equation}\label{App:eq:CD1}
    \mathbf{\dot{u}} = \mathbf{M_u} \mathbf{A_u} + \rho_u 
\end{equation}
where $\mathbf{\dot{u}} \in \mathbb{R}^L$ and we have the delay matrix $\mathbf{M_u} \in \mathbb{R}^{L \times 3}$.
To check if there is a causal connection from  $V$ to $U$, we append the delay matrix from the other time series, $\mathbf{M_v}$, to the equation
\begin{equation}\label{App:eq:CD2}
    \mathbf{\dot{u}} = (\mathbf{M_u} \mathbf{M_v}) \mathbf{D} + \rho_{uv}.
\end{equation}
Now $\mathbf{M_u} \mathbf{M_v} \in \mathbb{R}^{L \times 6}$, resulting in $\mathbf{D} = (d_1, d_2, \dots, d_6) \in \mathbb{R}^6$.
If there is a causal connection from  $V$ to $U$, then the last three elements of $\mathbf{D}$ will make the model better and the error $\rho_{uv}$ should decrease. 
If there is no causal connection from  $V$ to $U$ and the data windows are not dynamically similar, then the last three elements of $\mathbf{D}$
will be irrelevant and the error $\rho_{uv}$ should not change.
The difference
\begin{equation}
    \mathcal{C}_{uv} = \abs{\rho_u - \rho_{uv}}
\end{equation}
can therefore be used to quantify causality from $V$ to $U$.
A causal connection from  $U$ to $V$ can be tested in the same way, starting with
\begin{equation}\label{App:eq:CD3}
    \mathbf{\dot{v}} = \mathbf{M_v} \mathbf{A}_v + \rho_v
\end{equation}
where $\mathbf{\dot{v}}$ is a vector of length $L$ and $\mathbf{M_v} \in \mathbb{R}^{L \times 3}$.
Once again, the second delayed matrix $\mathbf{M_u}$ can be appended to the equation,
\begin{equation}\label{App:eq:CD4}
    \mathbf{\dot{v}} = (\mathbf{M_v} \mathbf{M_u}) \mathbf{G} + \rho_{vu}.
\end{equation}
We have $\mathbf{M_v} \mathbf{M_u} \in \mathbb{R}^{L \times 6}$ whence $\mathbf{G} \in \mathbb{R}^6$.
Whether the last three terms of $\mathbf{G}$ are relevant or not tells us whether there is a causal connection and 
\begin{equation}
    \mathcal{C}_{vu} = \abs{\rho_v - \rho_{vu}}
\end{equation}
is used to quantify causality from $U$ to $V$.  

However, this and all other causality measures assume that the two dynamical systems are not entirely similar or synchronized to each other. In the case of $u(t) = x_2(t)$ and $v(t)=x_3(t)$ from  systems R2 and R3 in \eqref{eq:RosEx1} the two time series $x_2(t)$ and $x_3(t)$ are 
dynamically similar. 
Therefore, the matrix $\mathbf{M_v}$ in \eqref{App:eq:CD2} is dynamically similar to the matrix $\mathbf{M_u}$ in \eqref{App:eq:CD1} and the matrix $\mathbf{M_u}$ in \eqref{App:eq:CD4} is dynamically similar the matrix $\mathbf{M_v}$ in \eqref{App:eq:CD3}.
We use the phrase `dynamically similar' because the attractors are similar, even though the time series themselves are not.
Therefore, $\mathcal{C}_{uv}$ and $\mathcal{C}_{vu}$ will increase because $\mathbf{M_u} \mathbf{M_v}$ in Eqs.~[\ref{App:eq:CD2}] and [\ref{App:eq:CD4}] contain more data than $\mathbf{M_u}$ alone in \eqref{App:eq:CD1} or $\mathbf{M_v}$ alone in \eqref{App:eq:CD3}.
A similar effect causes the other false positive $2 \rightarrow 7$: The matrices $\mathbf{M}_{x_2}$ and $\mathbf{M}_{x_3}$ 
are dynamically similar and therefore also $(\mathbf{M}_{x_2} \mathbf{M}_{x_7})$ and $(\mathbf{M}_{x_3} \mathbf{M}_{x_7})$
are dynamically similar.
Therefore, a false positive is detected.

\subsubsection{Indirect connections}

Indirect connections represent pathways between two nodes that are not directly linked but are connected through one or more intermediate nodes. 
Ding et al.\ \cite{ding06-437} point out that indirect connections might be impossible to identify.
In our network example we have such an indirect connection from $3 \rightarrow 1$. 
When looking at Fig.~\ref{fig:Motifs_numerical_1}(a), this connection is immediately identified as background and therefore it is not an issue for CD-DDA.


\subsection{Unconnected Network}

\begin{figure}[htp]
    \centering
    \includegraphics[width=0.5\columnwidth]{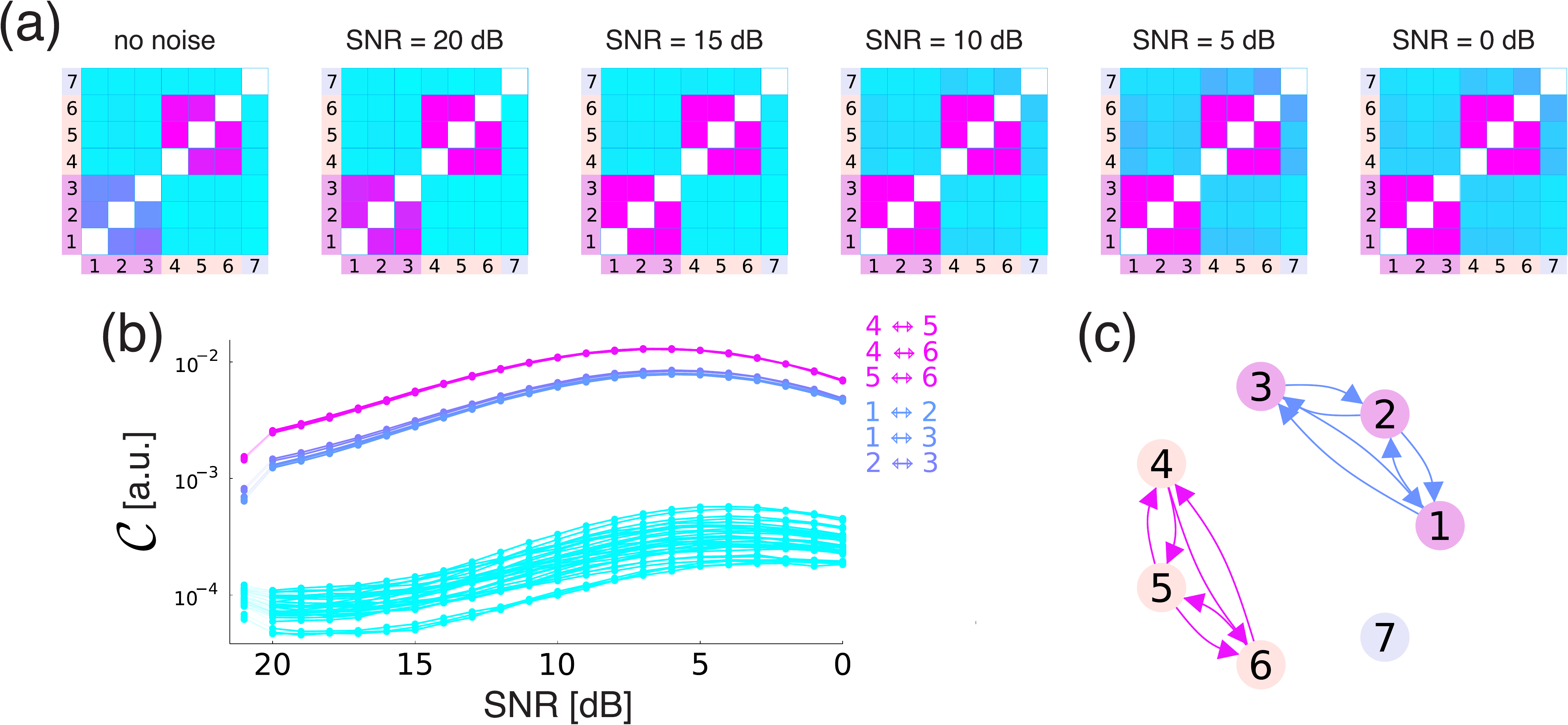}
    \caption{Seven independent R\"ossler systems with $\epsilon=0$ for all. (a) Numerically obtained adjacency matrices for varying SNRs under additive white noise, ~(b) dependence of ${\cal C}$ on SNR, ~(c) the numerical network motif with each edge corresponding to a false positive.}
    \label{fig:Motifs_numerical_0_i}
\end{figure}

\begin{figure}[htp]
    \centering
    \includegraphics[width=0.4\columnwidth]{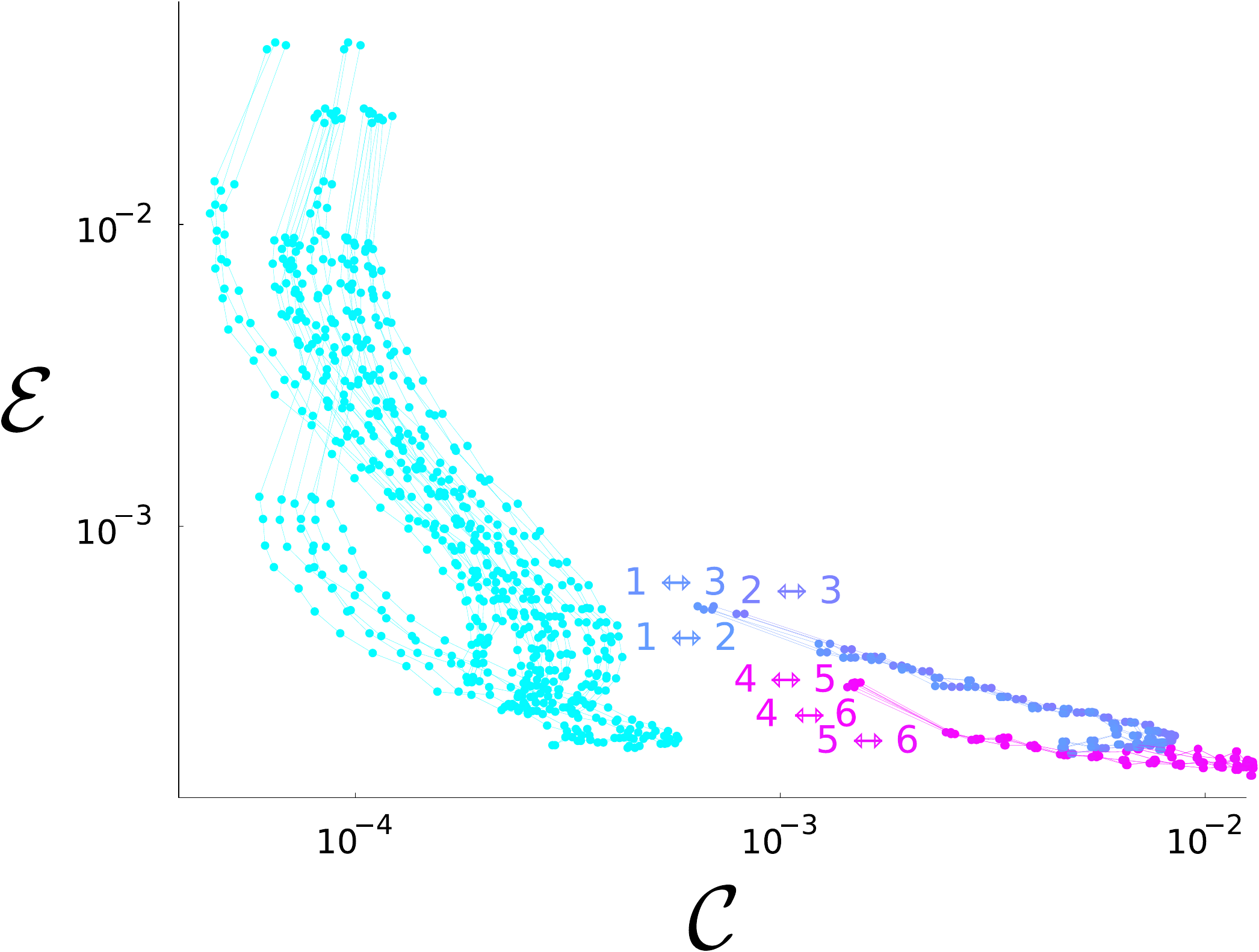}
    \caption{Dynamical ergodicity ${\cal E}$ over causality ${\cal C}$.
    The colors correspond to the ones in the numerically obtained adjacency matrix in \cref{fig:Motifs_numerical_0_i}(a) (left).
    The plot shows ${\cal E}$ over ${\cal C}$ for all nodes in the network with changing SNR.}
    \label{fig:C_E__SR__corrcoef__0_i}
\end{figure}

In this example all seven R\"ossler systems are independent from each other and all $\epsilon=0$ in \eqref{eq:ros}.
In~\cite{lainscsek21-103108} we identified the three dynamical states in the network using DE-DDA but could not identify those as false
positives due to their small magnitude. Here, they can be clearly identified as false positives. \Cref{fig:Motifs_numerical_0_i} shows the
numerically obtained adjacency matrices, the dependence of ${\cal C}$ on SNR, and the numerically obtained network motif. All edges are
false positives and show CDR in \cref{fig:Motifs_numerical_0_i}. \Cref{fig:C_E__SR__corrcoef__0_i} shows ${\cal E}$ as function of
${\cal C}$ and all edges can easily be identified as false positives.

\section{Epilepsy Data}
\label{sec:epilepsy}

In~\cite{lainscsek17-3181}, a genetic algorithm was used to select the model with minimum error from one second data segments for one
hour periods centered on the seizure onset times. Around one million such data segments  (155 seizures and 730 iEEG channels from 13
patients) were analyzed in this way.  The patient demographics and characteristics are described in~\cite{lainscsek17-3181}. All data
were obtained with informed patient consent and handled following protocols as approved by the IRB of the Massachusetts General Hospital.

The DDA model selected in~\cite{lainscsek17-3181} for the characterization of epileptic seizures is
 \begin{equation}\label{DDA_epilepsy}
     \dot{u} = a_1 u_1 + a_2 u_2 + a_3 u_1^4 
 \end{equation}
with $u_i = u(t-\tau_i)$
and $\tau_1=7, \tau_2=10$.
We used overlapping data windows of 250~ms length with a window shift of 25~ms. We did not filter or pre-process the data except z-scoring each data window.
Here, we show results from 2 hours of data (from half an hour before seizure \#4 to half an hour after seizure \#5) from one patient. 
We computed the DDA features $a_1, a_2, a_3$ and the error $\rho$ for each channel as well as ${\cal E}$ and ${\cal C}$ for each pairwise channel combination for the raw data as well as for the data with white noise added.  

The anatomical location of each electrode was identified by employing a combined volumetric and surface registration. We then mapped the channels to brain areas with an electrode labeling algorithm (ELA;~\cite{Felsenstein2019,Soper2023}), and grouped them into eight major regions indicated by the abbreviations in \cref{tab:brain-regions} and are preceded by R or L to indicate the right or left hemisphere.  For the patient shown here the regions LFP, LCG, LAT, LHP, RFP, RAT, RHP, and RSU had implants.

\begin{table}[htb]
    \centering
    \begin{tabular}{lc}
        \toprule
        Brain Region & Abbreviation \\
        \midrule
        frontoparietal & FP \\
        cingulate & CG \\ 
        lateral temporal & AT \\ 
        mesial temporal & MT \\ 
        hippocampus & HP \\ 
        occipital & OC \\ 
        thalamus & TH \\ 
        subcortical & SU \\
        \bottomrule
    \end{tabular}
    \caption{Brain regions with abbreviations.}
    \label{tab:brain-regions}
\end{table}

\begin{figure}
    \centering
    \subfigure[$\mathcal{C}$ for raw data. Numbers smaller than \(0.01\) are disregarded.]{\includegraphics[width=0.48\linewidth]{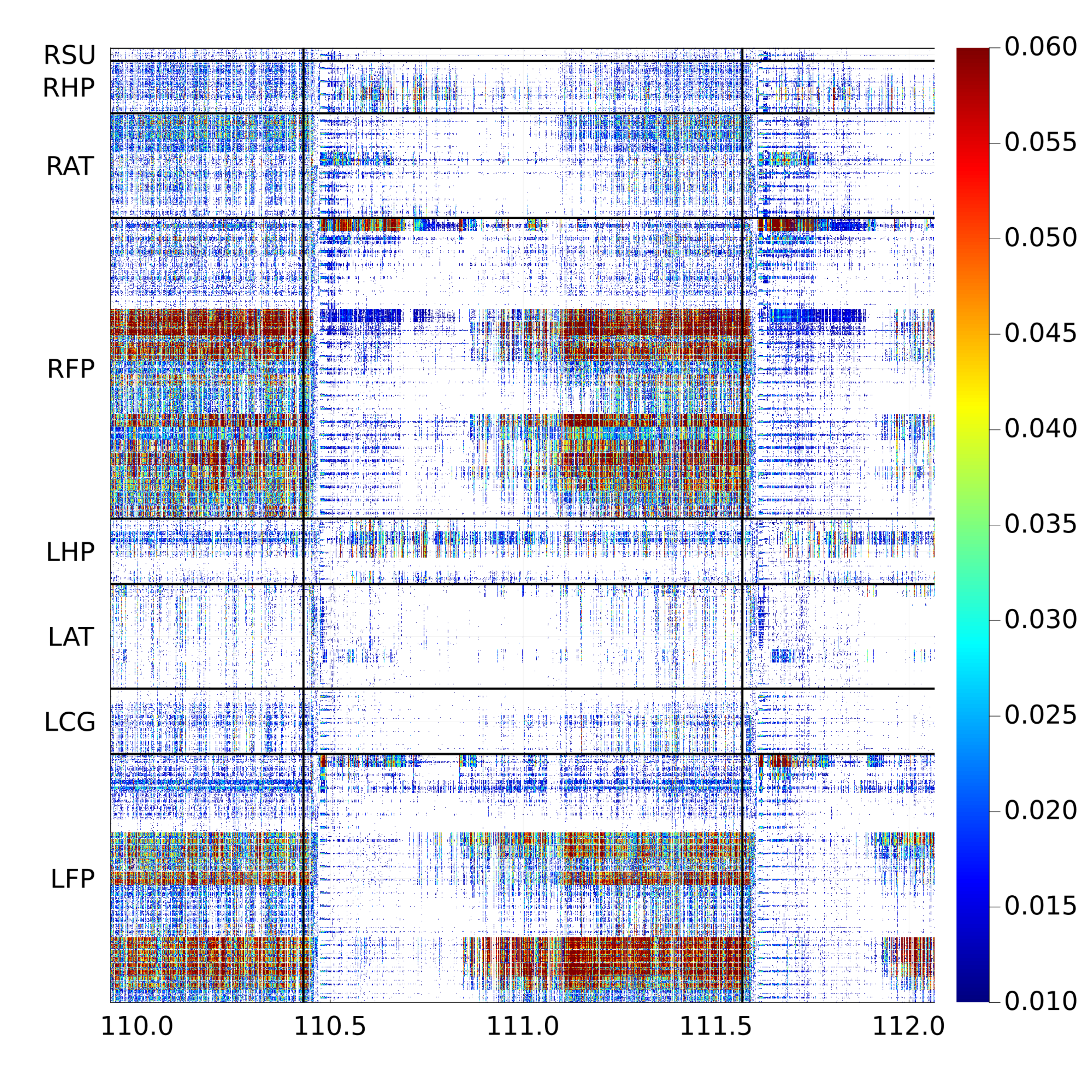}\label{fig:MG23__SZ4_SZ5__alpha_c1_c2_a}}%
    \subfigure[$\mathcal{C}$ for raw data without false positives.]{\includegraphics[width=0.48\linewidth]{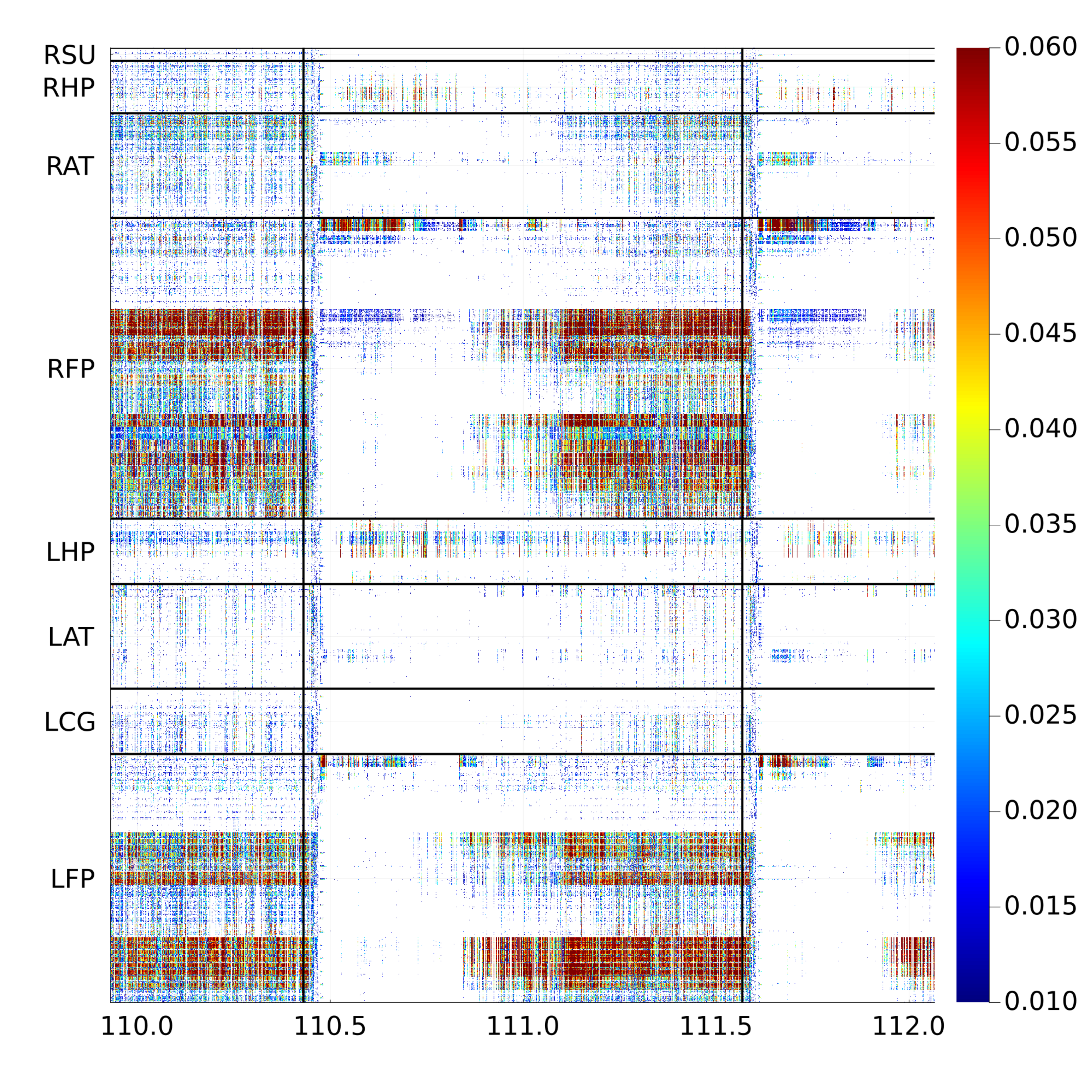}\label{fig:MG23__SZ4_SZ5__alpha_c1_c2_c}}
    \caption{Causality $\mathcal{C}$, or information flow, from the indicated brain regions listed above for two hours of data. All channels in each region are shown. The $x$-axes show hours since the beginning of the recording. The abbreviations for the brain regions on the $y$-axis are: The first letter indicates Left or Right. Followed by: FP:  frontoparietal, CG: cingulate, AT: lateral temporal, HP: hippocampus, and SU: subcortical.
    In (a) causality $\mathcal{C}$ is shown for all combinations of channels sorted by brain regions. 
    To test for true connections we added 15~dB white noise to the data. 
    For many connections $\mathcal{C}$ got stronger, which indicates causal dynamic resonance, or false positives. These were removed in (b). 
    The plot is much cleaner compared to (a) and the important connections can be seen more clearly.  
    }
    \label{fig:MG23__SZ4_SZ5__alpha_c1_c2}
\end{figure}

Fig.~\ref{fig:MG23__SZ4_SZ5__alpha_c1_c2} shows the temporal evolution of all pairwise connections between the channels, grouped according to the brain regions mentioned above. In Fig.~\ref{fig:MG23__SZ4_SZ5__alpha_c1_c2}(a), causality is shown for the raw data. We then added additional white noise with SNR~=~15~dB to the already noisy data. Connections with CDR were then removed as shown in Fig.~\ref{fig:MG23__SZ4_SZ5__alpha_c1_c2}(b).
This approach enables us to identify and retain the most meaningful connections, while filtering out those less significant to the system.

Identifying the dynamic network properties associated with seizures may improve both diagnosis and treatment. Visualization and analysis of highly connected nodes or hubs, together with interactions across distributed brain networks, offer deeper insight into network organization. These approaches enhance our understanding of seizure mechanisms from initiation to propagation and termination, and may help identify critical nodes or pathways that can be targeted for therapeutic intervention.

\section{Conclusion}%
\label{sec:conclusion}

We have introduced CDR to uncover spurious causal relationships within networks, thereby enhancing the accuracy of causality detection. CDR was rigorously studied using simulated data from seven coupled R\"ossler systems, where the known network structure enabled validation of its performance. We further applied CDR to invasive intracranial electroencephalographic (iEEG) recordings from patients with drug-resistant epilepsy during presurgical evaluation, demonstrating its applicability and potential utility in complex real-world neural data analyses.

We further outlined a possible similarity between CDR and Stochastic Resonance (SR). We must re-emphasize that these ideas are, while certainly titillating, an area of intense ongoing study, with a quantitative treatment well beyond the scope of the current paper.

\section{Code}

The code along with a detailed description is available
on the DDALAB website: \url{https://snl.salk.edu/~claudia/DDALAB/}.
This webpage includes links to Github and DDA related articles.
The computations from this paper are summarized on Github:  \\ \texttt{https://www.github.com/lclaudia/CDR}.

\section{Appendix}

\subsection{Causal Dynamic Recurrence Test}

In Fig.~\ref{fig:CD_test} we used the R\"ossler system
\begin{equation}
\begin{array}{rcl}
\dot{x} & = & - y - z \\
\dot{y} & = & x + 0.15 y \\
\dot{z} & = & b_i - 10 z + x z 
\end{array} 
\end{equation}
with $b_{1,2} = 0.2$ plus/minus a random number smaller than 0.005, $dt = 0.025$, a transient of 100000 data points, and random initial conditions. The integrated system resulted in 30000 data points and we ran DDA using [\ref{DDA-model}]
with a window length of 3000 data points and a window shift of 50 data points resulting in 540 data windows. \\
For the Lorenz system
\begin{equation}
\begin{array}{rcl}
\dot{x} & = & -10 x + 10 y  \\
\dot{y} & = & R_i x - y - x z \\
\dot{z} & = & - \frac{8}{3} z + x y
\end{array} 
\end{equation}
with $R_{1,2} = 28$ plus/minus a random number smaller than 0.005, $dt = 0.01$, a transient of 100000 data points, and random initial conditions. The integrated system resulted in 30000 data points and we ran DDA with a window length of 3000 data points and a window shift of 50 data points resulting in 540 data windows.
For this system we used the DDA model 
\begin{equation}
\dot{u} = a_1 u_1 + a_2 u_1^2 + a_3 u_1^2 u_2  
\end{equation}
with $u_j = u(t - \tau_j)$, $\tau_1 = 6~\delta t$,  $\tau_2 = 20~\delta t$, and $\delta t = 0.01$. 

\subsection{Modified R\"ossler System}

In \cite{lainscsek11-046205} we asked the question:  Is it possible that the time series corresponds to a function of the variables $\phi(x_i)$? For simplicity, does it correspond to one of the variables $x_i$? This question has been answered for the example of the  $x_2$-variable of the R\"ossler system \cite{roessler76-397}.
Here, we investigate the response system in Eq.~[\ref{eq:palus}]
\begin{equation}
\begin{array}{rcl}
\dot{x}_{2} & = & a_{1,1} x_2 + a_{1,2} y_2 + a_{1,3} z_2 + a_{0,0} x_1 
\\ & = &
-\epsilon x_2 -\omega_2 y_{2} - z_{2}  + \epsilon x_{1}
\\
\dot{y}_{2} & = & a_{2,1} x_2+ a_{2,2} y_2
\\ & = &
\omega_2 x_{2,j} + a y_{2} \\
\dot{z}_{2} & = & a_{3,0}+ a_{3,3} z_2 + a_{3,6} x_2 z_2
\\ & = &
b - c z_{2} + x_{2} z_{2} 
\end{array}
\end{equation}
in a similar way.
We start by assuming $y_2$ to be the  observed scalar signal
and is then  rewrite it as differential model starting from e.g.\  $X=y_2$ as
\begin{equation}\begin{array}{rcl}
 \dot{X} &=& Y \\
 \dot{Y} &=& Z \\
 \dot{Z} &=& F(X,Y,Z) \end{array}
\end{equation}
where the successive derivatives of $X=y_2$
define the new state space variables
$Y$ and $Z$. The function $F(X,Y,Z)$ is explicitly
\begin{equation}
\label{ros_DM}
\begin{array}{rcl}
F(X,Y,Z) & = &
\alpha _1 + X \alpha _2 + X^2 \alpha _3 + Y \alpha _4 + 
\\ & & 
X Y \alpha _5
+ Y^2 \alpha _6 + 
Z \alpha _7 + 
\\ & & 
X Z \alpha _8 + 
Y Z \alpha _9
\end{array}
\end{equation}
where the successive derivatives of $X=y_2$
define the new state space variables
$Y$ and $Z$. This functional form is the same as for the original R\"ossler system \cite{lainscsek11-046205}.
The coordinates $\alpha_r$ of the differential
embedding are related to the coordinates $a_{i,*}$ of the
modified R\"ossler model by
\begin{equation}
\label{ros_alpha}
\begin{array}{rcl}
\alpha_1 & = & a_{1,3}   a_{2,1}   a_{3,0} -a_{0,0} a_{2,1} a_{3,3}\\
\alpha_2 & = & -a_{1,2} a_{2,1}  a_{3,3} +a_{1,1} a_{2,2} a_{3,3}+a_{0,0} a_{2,2} a_{3,6} \\ [1ex]
\alpha_3 & = & a_{1,2}  a_{2,2}  a_{3,6} 
-\ffrac{a_{1,1} a_{2,2}^2 a_{3,6}}{a_{2,1}} \\
\alpha_4 & = & a_{1,2}  a_{2,1}-a_{2,2}  a_{3,3} 
\\ & &
-a_{1,1} ( a_{2,2} + a_{3,3})
-a_{0,0} a_{3,6} \\ [1ex]
\alpha_5 & = & \ffrac{a_{2,2}^2  a_{3,6}}{a_{2,1}} - a_{1,2}  a_{3,6} 
+\ffrac{2 a_{1,1} a_{3,6} a_{2,2}}{a_{2,1}} 
\\ [2ex]
\alpha_6 & = & -\ffrac{a_{2,2}  a_{3,6}}{a_{2,1}} 
-\ffrac{a_{1,1} a_{3,6}}{a_{2,1}} 
\\ [1ex]
\alpha_7 & = & a_{2,2}+a_{3,3} + a_{1,1}\\ [1ex]
\alpha_8 & = & -\ffrac{a_{2,2}  a_{3,6}}{a_{2,1}} \\
\alpha_9 & = & \ffrac{a_{3,6}}{a_{2,1}} \,\,.
\end{array}
\end{equation}
To uncover the interconnections between the parameters $a_{i,*}$ we a scaling  transformation can be used. It is typical that if some set of parameters
$a_{i,*}$ satisfies the inverse transformation [\ref{ros_alpha}], then
a new set of parameters $\tilde{a}_{i,*}$ also satisfies
the inverse transformation.  The new parameters are related to
the original set by a scaling transformation
$\tilde{a}_{i,*} = \lambda^{p(i,*)}a_{i,*}$.
The simplest way to determine these scaling
relations, specifically the set of exponents $p(i,*)$, 
is to note that each coefficient $\alpha_r$ is
a sum of products of powers of the original model
parameters $a_{i,*}$.  Take the logarithms of these
nonlinear product functions, construct the appropriate
coefficient matrix, and look for the null space.
Details can be found in \cite{lainscsek11-046205}.

We repeat the approach for $X = x_2$ and $X = z_2$ and 
combine these results to obtain Eq.~[\ref{eq:palusR}].
This means that any choice of $m=n$ and $k$ will generate exactly the same time
series of $y_2(t)$ and any choice of $n$ and $k$ will generate exactly the same time
series of $z_2(t)$.
It also means that the parameter $a_{1,1}$ in the term $a_{1,1} x_2 = - \epsilon x_2$ is an additional bifurcation parameter of the modified R\"ossler system while the term $a_{0,0} x_1 = \epsilon x_1$, that comes from the driver system, is connected to all the parameters of the modified R\"ossler system except $a$, $c$, and $a_{1,1}$.

\section*{Acknowledgements}
The authors thank Mark Spano, Eduardo M.\ A.\ M.\ Mendes, and Vin\'icius Rezende Carvalho  for valuable discussions.
C.L. and T.S. acknowledge support by the grants 
NIH MH132664 and ONR N00014-23-1-2069.
P.S. acknowledges support by the Department of Defense CDMRP FY21 Epilepsy Research Program W81XWH-22-1-0315.


\end{document}